\pdfoutput=1
\documentclass[journal=jacsat,manuscript=article]{achemso}
\usepackage[version=3]{mhchem} 
\usepackage{adjustbox}
\usepackage{lscape}
\usepackage{textcomp}

\usepackage{lineno}

\author{Giulia Ceriotti}
\affiliation{Institute of Earth Surface Dynamics, University of Lausanne, Switzerland}
\email{giulia.ceriotti@unil.ch}
\author{Sergey M. Borisov}
\affiliation{Institute of Analytical Chemistry and Food Chemistry, Graz University of Technology, Austria}
\author{Jasmine S. Berg}
\affiliation{Institute of Earth Surface Dynamics, University of Lausanne, Switzerland}
\author{Pietro de Anna}
\affiliation{Institute of Earth Sciences, University of Lausanne, Switzerland}
\email{pietro.deanna@unil.ch} 
\phone{+41 21 692 4423}

\title{Morphology and size of bacterial colonies control anoxic micro-environments formation in porous media}

\abbreviations{IR,NMR,UV}

\begin{document}
	
%
%
%
%

\begin{abstract}
Bacterial metabolisms using electron acceptors other than O$_2$ (e.g., methanogenesis and fermentation) largely contribute to element cycling and natural contaminant attenuation/mobilization, even in well-oxygenated porous environments, such as shallow aquifers. This paradox is commonly explained by the occurrence of small-scale anoxic micro-environments generated by the coupling of bacterial respiration and the dissolved oxygen (O$_2$) transport by pore-water. Such micro-environments allow facultative anaerobic bacteria to proliferate in oxic environments. Micro-environments dynamics are still poorly understood due to the challenge of directly observing biomass and O$_2$ distributions at the micro-scale within an opaque sediment matrix. To overcome these limitations, we integrated a microfluidic device with transparent O$_2$ planar optical sensors to measure the temporal behavior of dissolved O$_2$ concentrations and biomass distributions with time-lapse video-microscopy. Our results reveal that bacterial colony morphology, which is highly variable in flowing porous systems, controls the formation of anoxic micro-environments. We rationalize our observations through a colony-scale Damköhler number comparing dissolved O$_2$ diffusion and bacterial O$_2$ uptake rate. Our Damkholer number enables us to predict the pore space fraction occupied by anoxic micro-environments in our system for a given the bacterial organization.
\end{abstract}

\noindent
\textbf{Synopsis}: Sediments and soils are heterogeneous environments populated by active bacteria. This study explores the coupling of aerobic bacteria respiration and oxygen transport as the key driver for the formation of pore-scale anoxic micro-environments.

\section{Introduction}

Although the Earth's atmosphere is well-oxygenated, porous subsurface environments fully or partially saturated by water (such as aquifers, lake/ocean sediments, and soils) are often oxygen (O$_2$)-limited~\cite{berg2022low}. Such limitation arises from the low O$_2$ solubility in water and the porous solid matrix of sediments and soils, constituted by a disordered arrangement of grains, that physically hinders the O$_2$ diffusion and supply from the external atmosphere~\cite{or2007physical,pedersen2015measuring,keiluweit2018anoxic}. As a result, O$_2$ availability typically decreases with depth, in the subsurface.~\cite{haberer2011high,parker2014dissolved}\\

\noindent
Since O$_2$ is the most thermodynamically favorable common electron acceptor, its spatial distribution in the subsurface remarkably influences the activity of the diverse bacteria that populate these natural environments~\cite{keiluweit2018anoxic,or2007physical}. Aerobic respiration is associated with the shallow subsurface that is generally well-oxygenated. Processes using less thermodynamically favorable electron acceptors than O$_2$ (such as nitrates, iron-oxides, sulfates, and CO$_2$) are typically confined to deeper and reducing environments~\cite{herath2016natural,boschetti2008springs,keiluweit2018anoxic}. Surprisingly, recent investigations have shown the occurrence of non-aerobic metabolisms in well-oxygenated regions of the subsurface~\cite{kravchenko2017hotspots,kuzyakov2015microbial,rubol20162d,schramm1999occurrence}. For example, oxygenated and well-drained superficial soils largely contribute to methane and NO\textsubscript{X} emissions generated by facultative methanotrophs~\cite{keiluweit2017anaerobic} and denitrifiers~\cite{kravchenko2017hotspots}. Also, Mn reduction and reductive dehalogenation are active processes in oxic aquifers and are crucial players in the natural attenuation and bio-remediation of recalcitrant pollutants that accumulate in groundwater~\cite{herrero2022key,visser2021fate}.\\   

\noindent 
This apparent paradox results from our limited capacity to fully characterize the pore-scale heterogeneity of the geo-chemical environment experienced by bacteria in the subsurface~\cite{baveye2018emergent}. Depending on the sampling technique, bulk O$_2$ measurements typically capture the average oxygenation level over the centimeter-meter scale~\cite{parker2014dissolved}. Although the bulk system appears well-oxygenated at that scale, pore water may display extremely heterogeneous O$_2$ distributions over smaller spatial scales (10$\sim$1000~$\mu$m)~\cite{or2007physical} that bacteria directly experience and to which they actively respond~\cite{smriga2021denitrifying,de2021chemotaxis}. At such small scales, diffusion should be very efficient in smoothing O$_2$ gradients but the tortuous architecture of soils and sediments hinders homogenization by diffusion~\cite{kuzyakov2015microbial,smriga2021denitrifying,zhang2010effects,or2007physical,Hamada2020}. Pore-water flow (for instance in aquifers and flooded soils), being typically organized in preferential flowing paths and stagnant zones, further contributes to the O$_2$ heterogeneity~\cite{zhang2010effects,kuzyakov2015microbial}. Bacteria grow most abundantly along the O$_2$ and nutrient-rich flow paths~\cite{zhang2010effects,franklin2019more} where they form highly active hotspots~\cite{kravchenko2017hotspots}. Shear stresses generated by flow and porous geometry~\cite{hommel2018porosity,carrel2018biofilms}, in addition to nutrient concentrations and bacterial growth rate~\cite{young2022pinning}, constrain colony sizes and morphology (as thin surface layers, round clusters, clumps, or filaments) which influence colony growth rates and, consequently, O$_2$ uptake~\cite{tronnolone2018diffusion,vulin2014growing}.\\

\noindent
The formation of local micro-scale zones of low oxygen level, undetected by bulk measurements, may provide micro-habitats for facultative anaerobic bacteria that switch to alternative metabolic pathways other than O$_2$ respiration\cite{baveye2018emergent}. For different facultative strains, the complete depletion of free O$_2$ in the environment is not a prerequisite to respire electron acceptors other than O$_2$, implying that aerobic respiration and alternative facultative metabolisms can occur simultaneously and synergically in these low oxygen level zones\cite{marchant2017denitrifying,pedraz2020gradual}. Going beyond the strict definition of anoxic as the complete absence of free O$_2$, we define the local micro-scale low oxygen areas as anoxic micro-environments. These are the zones where facultative anaerobic bacteria might use less thermodynamically favorable electron donors than O$_2$ for metabolic purposes. These zones are characterized by levels of oxygen that significantly differ from bulk measurements. \\


\noindent 
Anoxic micro-environments have been observed to occupy no more than 10\% of the pore space, are generally smaller than a few millimeters, and persist from hours to days~\cite{kuzyakov2015microbial,keiluweit2018anoxic,groffman2009challenges}.
Despite the small portion of pore volume occupied, anoxic micro-environments are nowadays considered fundamental to explain the dynamics of many macro-scale ecological processes such as element cycling (e.g., soil carbon stabilization~\cite{keiluweit2017anaerobic} and sulfur reduction and precipitation~\cite{raven2021microbial}), greenhouse gas production (e.g., methanogenesis~\cite{ye2019methane,bivzicmethane} and NO\textsubscript{x} emissions~\cite{cook2017does,groffman2009challenges,smriga2021denitrifying}), heavy metal mobilization~\cite{lehto2017mesocosm,widerlund2007size} and natural attenuation of recalcitrant pollutants that accumulate in groundwater~\cite{herrero2022key,visser2021fate}.\\

\noindent
Yet, a quantitative understanding and predictability of anoxic micro-environments formation in oxic subsurface porous environments is missing due to major methodological limitations. Indeed, direct in-situ monitoring of biomass growth and O$_2$ distribution at the colony scale remains challenging due to i)~the opacity of porous matrices of soils and sediments; and ii)~the difficulties in simultaneously capturing the different spatial and temporal scales of interest for anoxic micro-environments formation~\cite{baveye2018emergent,pedersen2015measuring,kravchenko2017hotspots}. 
The recent development of luminescent planar sensors~\cite{borisov2011novel,pedersen2015measuring,sun2015imaging}, combined with microscopy, has opened new perspectives enabling the investigation of O$_2$ spatial distribution at micro-scale resolution. The study of anoxic micro-environments with planar sensors has nevertheless been limited by the use of opaque sensor foils, which hinder visualization of biomass spatial organization and growth measurements~\cite{rubol20162d,keiluweit2018anoxic,lehto2017mesocosm,franklin2019more}. \\

\noindent
To overcome these limitations, we developed an experimental setup that combines transparent O$_2$ planar sensors with microfluidic devices that reproduce the laminar pore water flow regime through a porous medium, mimicking sandy sediments~\cite{aleklett2018build}. With time-lapse automated video-microscopy, we simultaneously monitor bacterial colony growth and O$_2$ concentrations across the entire porous landscape (covering tens of millimeters with micron resolution) over time, encompassing more than a hundred bacterial generations. Our experimental results provide evidence of the critical role played by the bacterial growth rate and colony morphology in controlling micro-scale O$_2$ heterogeneity in confined porous systems under laminar flow conditions. To rationalize our observations, we introduce a colony-specific Damköhler number. The latter is computed by comparing the biomass O$_2$ consumption rate and the O$_2$ diffusion rate within each individual colony. The so-defined Damköhler number, accounting for bacterial cluster's morphology, can identify the bacterial colonies prone to trigger anoxic micro-environments.\\

\section{Materials and Methods}
\label{sec: Materials}

\textbf{Bacterial cultures}.\\
\textit{Pseudomonas putida} sp. GB1 wild type is a non-motile aerobic soil-dwelling bacterium. Inoculum from frozen stock was grown overnight in 5~mL liquid Luria Bertani (LB) medium at 30\textdegree C in an orbital shaker set at 150 rpm. A  $100~\mu$L aliquot was resuspended in 5~mL of LB medium and incubated under the same conditions until the early exponential growth stage (i.e., for about 2~hours).\\

\noindent
\textbf{Integration of O$_2$ planar sensors within porous structures}.\\ We used customized transparent optical sensors to monitor spatial oxygen distribution within the  microfluidics device. These sensors rely on a mixture of two luminescent dyes immobilized into a polymeric matrix (polystyrene). The emission of one of the two dyes (Pt(II)~porphyrin) is quenched by molecular oxygen whereas the second dye provides a reference signal (see sections S1 and S2 in SI for details).  The planar sensor is prepared from a solution of the two dyes, polymer in organic solvent (anisole), that was screen-printed onto a glass slide (25~x~75~mm) and, after solvent evaporation, resulted in a thin layer (below~5~µm) that covered an area of 3~x~27~mm\textsuperscript{2}. The glass slide supporting this O$_2$ sensor was integrated into a fully transparent microfluidic device as described in the following.\\

\begin{figure}
	\centering
	\includegraphics[width=1\textwidth]{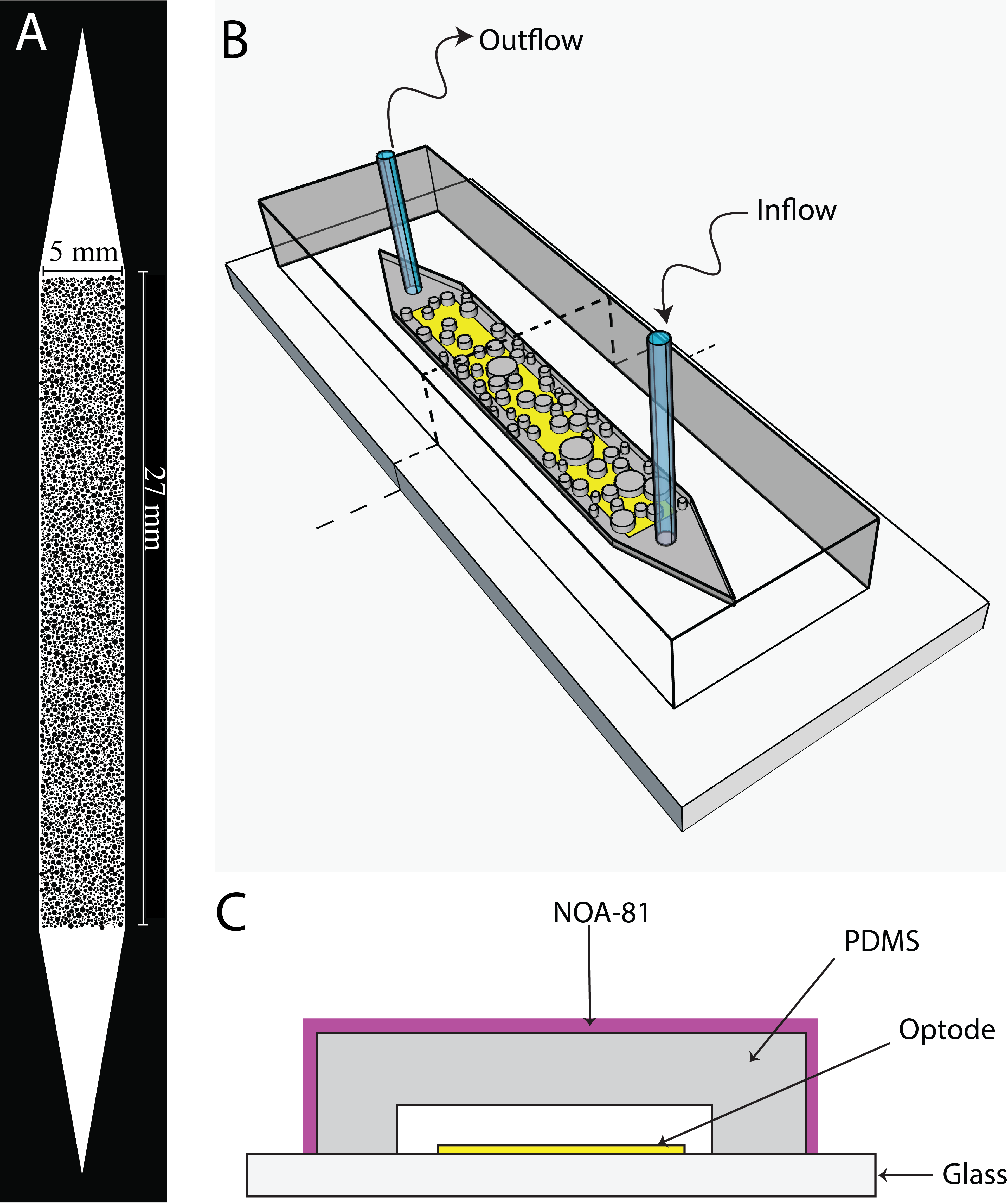}
	\caption{(A) Porous medium geometry used in the experiments. (B) Outline of the microfluidics device composed of glass slide for microscopy, transparent O$_2$ planar sensor, the PDMS chip engraving the porous medium geometry. Two tubes are connected the pore space of the chip to impose the desired flow conditions. (C) Cross-section of the microfluidics device (dashed black line in panel B) highlighting the layered structure of the chip.}
	\label{fig1}
\end{figure}

\noindent
We designed a two-dimensional porous medium geometry (Figure~\ref{fig1}A) composed of non-overlapping disks with varying diameters and centered in random locations. The minimum distance between neighboring grains $l_{pt}$ (referred to as pore-throat) is distributed according to a power law probability density function  ($p(l_{pt})~\sim ~l_{pt}^{-0.08}$). The average pore size is equal to 0.05~mm)~\cite{de2017prediction}. The resulting velocity field in porous space is also power law distributed~\cite{scheidweiler2020trait} and spans five orders of magnitude, consistently with those observed in natural sediments, e.g., sandstone~\cite{ceriotti2019double,alhashmi2016impact}.\\

\noindent
Using classical soft lithography, the porous geometry (Figure~\ref{fig1}A) was printed onto a microfluidics master with a thickness of $h~=~50~\mu$m (i.e., similar to the average pore-throat size~$l_{pt}$). This was used to mold PDMS chip replicates (Sylgard 184 Silicone Elastomer mixed with 10 w/w~\% of curing agent; supplier: Dow Corning, Midland, MI). 
The microfluidic device was assembled by aligning and plasma bonding the PDMS chip to a glass slide previously covered by the optodes (Figure~\ref{fig1}B) so that the sensor was in direct contact with the fluid occupying the porous space. The resulting porous system had a length $L~=~27$~mm, width $W~=~5$~mm, thickness $h$ and porosity $\phi~=~0.5$. \\
Three replicates were prepared by coating the chip's external surface with gas-impermeable NOA-81 glue (Figure~\ref{fig1}C) before the experiment. This prevents O$_2$ diffusion through the PDMS walls from the atmosphere and limits O$_2$ supply to the inlet flow. Two controls without NOA-81 coating were performed for comparison.\\

\noindent
\textbf{Flow conditions}.\\
Tygon tubing (Cole-Palmer, inner/outer diameter $0.02/0.04$ inches) was used to connect the two ends of the microfluidics device (Figure~\ref{fig1}B). After degassing, the chip was saturated by injecting the prepared cell suspension through the outlet tube to avoid cell accumulation close to the inlet. After a rest phase of 30 minutes, the inlet tube was replaced with a clean one prior to the injection of a well-oxygenated sterile and nutrient-rich solution (a 10X diluted Luria Bertani broth) for 45 hours at a flow rate of $0.2~\mu$L/min. The resulting Darcy velocity $q=Q/(\phi A)$ was equal to $0.013~$mm/s and the Reynolds number (assuming the characteristic length $l$ equal to the pore throat average size $l_{lp}~=~0.05~$mm, and kinematic viscosity $\nu~=~1$~mm\textsuperscript{2}/s) was $Re~=~q~l/~\nu~\sim~10^{-3}$, corresponding to a laminar flow regime. The P\'eclet number associated with the O$_2$ transport is $Pe~=~q~l/~D_m~=~0.3$ (where the O$_2$ diffusion coefficient is $D_m~=~2\cdot10^{-3}$~mm$^2$/s) and it indicates a diffusion-limited O$_2$ transport. The characteristic time necessary for the average flow to displace a fluid volume equivalent to the whole porous system is $t_{PV}~=~L/q~=~33~$~min. \\

\noindent
\textbf{Data acquisition}.\\A fully-automated microscope (inverted Nikon Eclipse Ti-E2) equipped with a 10X objective and an sCMOS camera (Teledyne Phometrics Prime 95B, sensor area $13.2\times~13.2$~mm) captured images of the entire porous landscape in three different optical configurations with a resolution of 1.1~$\mu$m/pixel every $\Delta~t~=~1$~hour. \\
The first optical configuration is an adjusted phase-contrast~\cite{de2021chemotaxis} to monitor the biomass distribution within the chip. The second and third optical configurations capture, separately, the luminescent signals of the two dyes composing the optodes using custom-made filter cubes (Semrock single bandpass ($500~\pm~20$~nm) and Semrock single-band pass filter ($662\pm11$)~nm). The Lumencor SPECTRA-X LED Light Engine, irradiating at 450~nm, was used for dye excitation.\\
Large images were obtained by stitching together 40 gray-scale pictures (16-bit) into a single image of $24,522~\times~2,814$~pixels (equivalent to $26.97$~mm by $3.09$~mm), capturing the entire planar sensor surface.\\

	\begin{figure}
		\centering
		\includegraphics[width=\textwidth]{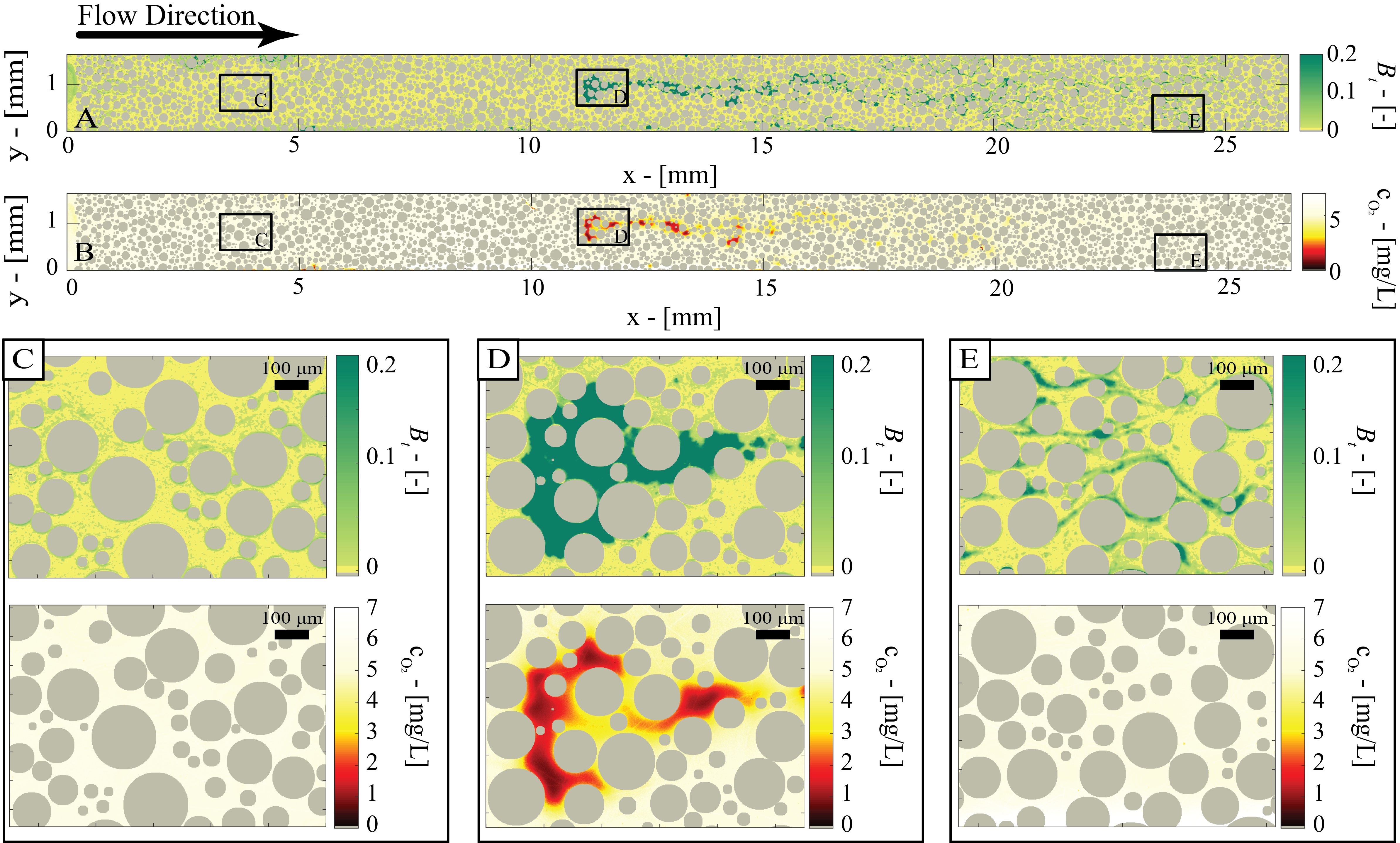}
		\caption{Spatial maps of the biomass $B_t$  (A) and O$_2$ concentration $c_{O_2}$ (B) at 47~$t_{PV}$ for limited O$_2$ supply (NOA-81~coated~chip). (C-E) Zoom-ins  of portions of O$_2$ map and biomass distribution presented in panel A and B delimited by solid black lines. Grey disks represent solid grains.}
		\label{fig2}
	\end{figure}


\noindent 
\textbf{Image processing and bulk quantities}.\\
The images collected during each experiment were post-processed to describe porous geometry, spatial and temporal maps of biomass and O$_2$ concentration at the pore scale. The porous geometry was obtained from an image taken before the inoculation with the adjusted phase-contrast configuration and stored as a binary matrix $M$ with zero value assigned to solid grains, one otherwise (see further details in the SI, section S3). Hereafter, the solid grains are displayed in all the spatial maps as gray disks which are systematically removed from images analyses (e.g., Figure~\ref{fig2}).\\

\noindent
The maps of biomass distribution over time, $B_t$, were obtained by identifying differences in the light intensity between the adjusted phase-contrast image at time $t$ against the image at $t~=~0$. Even though we did not directly correlate the $B_t$ maps to well-established metrics for biomass density (e.g., carbon density or Optical Density), we used the dynamics of $B_t$ as a proxy for biomass growth. The concentration $c_{O_2}$ of dissolved oxygen was obtained by dividing the O$_2$-sensitive optode signal by the reference one and converting this ratio into O$_2$ concentration values via a proper calibration procedure (described in the SI, section S2). 
The average biomass growth and O$_2$ concentration in the system are defined as $E[B_t]$ and $E[O_2]$, computed as:
\begin{equation}\label{eqn:EB}
	E[B_t](t) = \frac{\sum_{i} B_t(i,t)}{\sum_{i} M(i)} \quad \text{and} \quad E[c_{O_2}](t) = \frac{\sum_{i} c_{O_2}(i,t)}{\sum_{i} M(i)}
\end{equation}
%
\noindent
where $i$ is the image pixel counter and $M$ the aforesaid geometry binary matrix, which is time-independent.

\section{Results and Discussion}

\begin{figure}
	\centering
	\includegraphics[width=\textwidth]{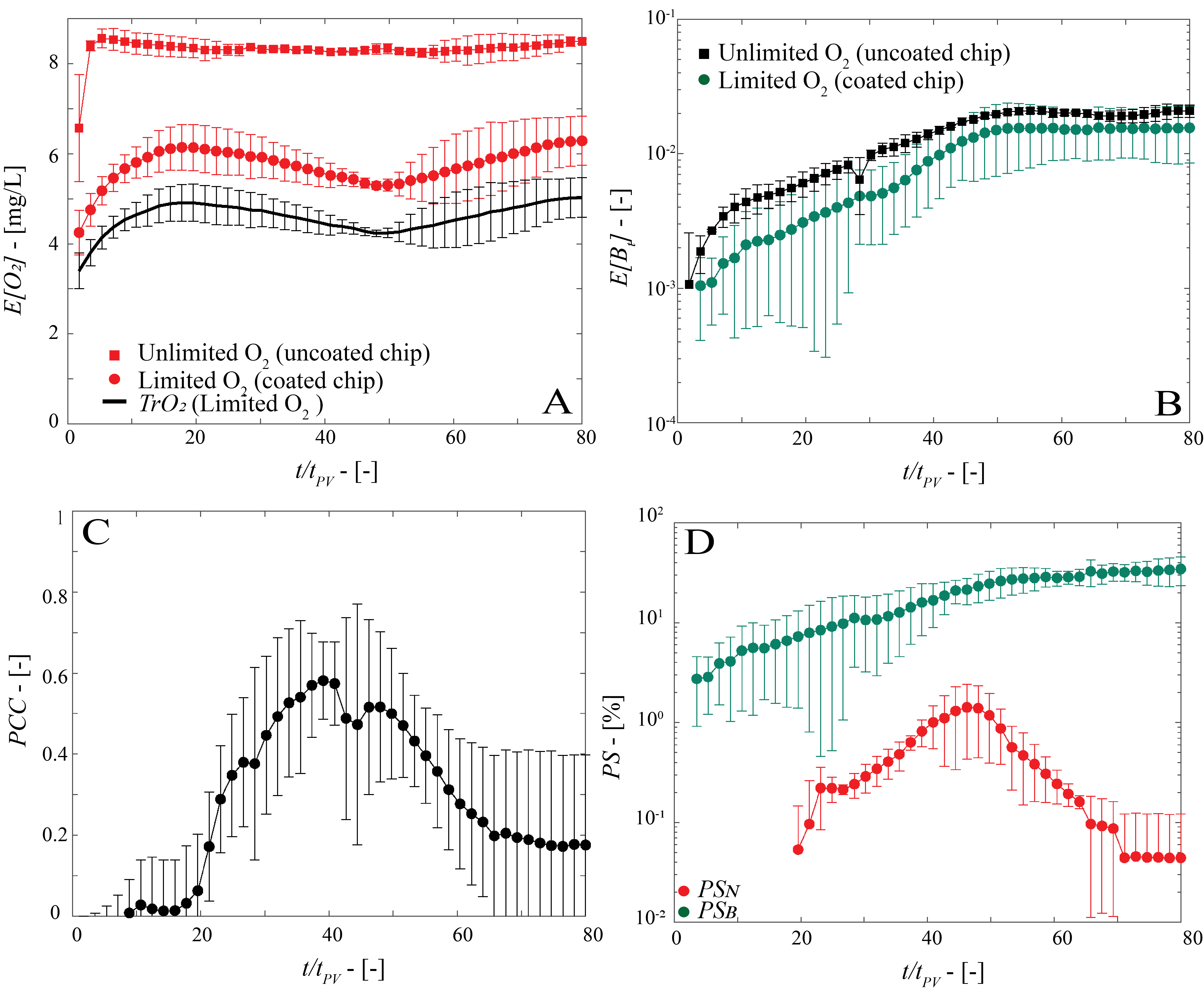}
	\caption{Temporal behavior of bulk quantities $E[O_2]$ (A) and $E[B_t]$ (B), computed according to  Eq.~\ref{eqn:EB}, for coated (circles) and uncoated (squares) systems. The solid black line in panel A indicates the thresholds ($TrO_2$) used to identify the anoxic micro-environments in the limited O$_2$ supply experiment (NOA-81~coated~chip). (C) Temporal behavior of the adjusted Pearson Correlation Coefficient (PCC, mathematical definition is detailed in SI, section S4) for the coated system. (D) Temporal behavior of the pore space percentage occupied by anoxic micro-environments ($PS_N$) and by biomass ($PS_B$) for the coated system. Symbols indicate averaged values and the errorbars denote the associated standard deviation of replicates.}
	\label{fig3}
\end{figure}

\noindent
\textbf{Bulk measurements do not capture micro-scale biogeochemical heterogeneity}.\\ 
The temporal behavior of the bulk O$_2$ concentration in our porous systems, $E[O_2]$ (Figure~\ref{fig3}A), depends on the extent and distribution of O$_2$ supply in microfluidic device. On the one hand, when O$_2$ was allowed to freely diffuse thought the PDMS walls of the uncoated chips (used as control), the bulk dissolved O$_2$ concentration was constant over time and in equilibrium with the atmosphere (about $8.2$~mg/L). On the other hand, when the oxygen supply was limited to the inlet flow (by chip coating with transparent NOA-81 glue), the bulk O$_2$ concentration decreased to 5.5-6.5~mg/L. The temporal behavior of $E[O_2]$ is mirrored by the quantity $Tr_{O_2}$ (solid black line in Figure~\ref{fig3}A), which represents the chosen O$_2$ concentration threshold for the identification of the anoxic micro-environments and it will be deeply discussed in the following section. \\

\noindent
The O$_2$ supplied by the constant fluid injection was sufficiently high in both scenarios to not limit the overall biomass growth, quantified by $E[B_t]$ (Figure~\ref{fig3}B). The latter exhibits the same logistic trend~\cite{madigan2008brock} for the coated and uncoated scenarios. The exponential growth lasts until 50 pore volumes have been injected ($1t_{PV}~=~33~min$) and it was followed by a stationary phase (for $t>50~t_{PV}$) possibly indicating that the system reached its carrying capacity.\\ 
Bulk O$_2$ (Figure~\ref{fig3}A) decreased during exponential growth when bacterial O$_2$ consumption was high and increased again during stationary phase when the consumption rate slowed down. The overall impact of bacterial activity on bulk O$_2$ was mild and the system can still be considered highly oxygenated for the entire experiment duration~\cite{keiluweit2018anoxic}. Then, both scenarios (coated and uncoated chips) can be considered macroscopically well-oxygenated. \\

\noindent
Though bulk O$_2$ concentrations in our coated and uncoated systems remained relatively high and sustained similar biomass growth, micro-scale O$_2$ distribution differed remarkably (Figure 2A for the coated system and Figures S.10 and S.11 for the uncoated one in SI, section S8). In the gas permeable (uncoated) system, O$_2$ concentrations were homogeneous while in the impermeable (coated) system anoxic micro-environments formed throughout the porous landscape (Figure~\ref{fig2}A). This result provides evidence that low-oxygen spots may form and persist in well-oxygenated porous systems such as groundwater, saturated soils during intense rain events, and river sediments. Furthermore, the anoxic zones persist for several hours during the stationary phase and are still detectable after $55~t_{PV}$ (i.e., after 5 hours from the beginning of the stationary phase; see SI, section S6). This evidence suggests that bacterial colonies, even though not expanding their size, are still up-taking O$_2$. Otherwise, such small-scale gradients would have been smoothed out by diffusion within a time scale of a few seconds. Then, bulk monitoring does not capture the persistence of anoxic micro-environments and overlooks heterogeneity at the scale of interest for bacteria. \\
       
\noindent
\textbf{Anoxic micro-environments are correlated to bacterial colony aerobic respiration in space and time.}\\
\noindent
The spatial correlation between O$_2$-depleted zones and biomass density is quantified through an adjusted Pearson Correlation Coefficient (PCC in Figure~\ref{fig3}C, details on the mathematical definition in the SI, section S4). A high positive spatial correlation is observed from 20$t_{PV}$ (with a maximum of~PCC$\sim$~0.6), which includes the both the exponential and the following stationary phases of the system. This suggests that the spatial organization of bacterial colonies controls the formation of localized O$_2$-depleted zones, or anoxic micro-environments.\\ 

\noindent
Identifying anoxic micro-environments is often performed by fixing an absolute O$_2$ concentration threshold that still does not have an established universal value~\cite{berg2022low}. Alternatively, one can define anoxic micro-environments as those zones where facultative anaerobic bacteria use electron acceptors other than O$_2$. Even though conceptually plain, the identification of anoxic micro-environments remains not trivial. Indeed, each facultative strain realistically switches its metabolism at different O$_2$ concentrations~\cite{marchant2017denitrifying,angel2011activation}. In this framework, the O$_2$ concentration threshold for anoxic micro-environments identification is inevitably strain- and environment-specific. In this work, we do not investigate any metabolism switch and choose the threshold $Tr_{O_2}$  from the statistical analysis of the O$_2$ concentration maps (details in the SI, section S5). Base on this analysis, we defined anoxic micro-environments as the portions of the pore space where O$_2$ concentration (c$_{O_2}$) is at least~20\% lower than the bulk $E[O_2]$ measurements. As a consequence of its definition, the threshold $Tr_{O_2}$ ($\sim 0.8~E[O_2]$ shown in Figure 3A) is not a constant absolute value but rather system-specific and time-dependent.\\

\noindent
We used $TrO_2$ to compute the percentage of pore space occupied by anoxic micro-environments $PS_{N}$ (i.e., the pore space fraction where $c_{O_2}<TrO_2$, see SI, section S3) at each time $t$ (Figure~\ref{fig3}D).
In addition to the correlation with biomass distribution, anoxic micro-environments formation exhibits a temporal dependence on biomass growth rate. For $t<50t_{PV}$, anoxic micro-environments progressively increase in volume mirroring biomass exponential growth (Figure~\ref{fig3}B). During the stationary phase ($t~>50t_{PV}$), O$_2$ distribution becomes progressively more homogeneous and the anoxic micro-environments disappear (see SI, section S5, for spatial maps at different times) reflecting the slowdown of bacterial O$_2$ consumption.\\
The maximum volume of anoxic micro-environments was attained at 46$t_{PV}$ accounting for 1~-~2\% of the pore space, which is consistent with previous estimates of microbial hotspot occurrence~\cite{kravchenko2017hotspots}. This implies that the proportion of space occupied by anoxic micro-environments remains always considerably smaller than the one occupied by biomass at all times as shown by $PS_B$ (Figure~\ref{fig3}C) that is defined as the percentage of pore space where $B_t>0$ (see SI for details, section S3). \\    

\noindent
Our findings indicate that biomass growth is a necessary condition and a key controlling factor for anoxic micro-environments formation but, alone, it does not explain the occurrence of micro-environments (see e.g., Figures~\ref{fig2}C and E). \\  

\noindent
\textbf{Round shape of colonies favors the anoxic micro-environments formation}.\\
We observe that colony morphology significantly deviates from rounded shapes in our flowing porous system (Figure~\ref{fig2}A), which confirms previous findings~\cite{hommel2018porosity}. Colonies that grow radially are often associated with remarkable localized O$_2$ depletion (Figure~\ref{fig2}D). In contrast, elongated structures along flow direction do not affect local O$_2$ content (Figure~\ref{fig2}E). This work does not focus on processes governing colony morphology, which are investigated elsewhere~\cite{carrel2018biofilms}. Instead, we draw the attention to the role of colony morphology in controlling anoxic micro-environments formation.\\

\noindent  
 To characterize colony morphology, each cluster of connected pixels colonized by biomass was identified in biomass spatial maps~($B_t$) and labeled as a distinct bacterial colony. All the colonies were described in terms of planar area~$A$, perimeter~$P$, and the equivalent radius (defined as $R_{EQ}~=~\sqrt{A/\pi}$). The latter is representative of the characteristic length scale of a round colony. In addition to these metrics, we introduce a colony characteristic length scale, alternative to $R_{EQ}$, defined as $\lambda~=~2A/P$ which  discriminates between colonies with similar areas but different shapes. This can also be interpreted as the volume-to-surface ratio of 3D colonies adapted to our 2D images. If a colony is rounded, $\lambda$ approaches $R_{EQ}$ whereas for elongated colonies $\lambda$ attains values smaller than $R_{EQ}$. 

\begin{figure}
	\centering
	\includegraphics[width=\textwidth]{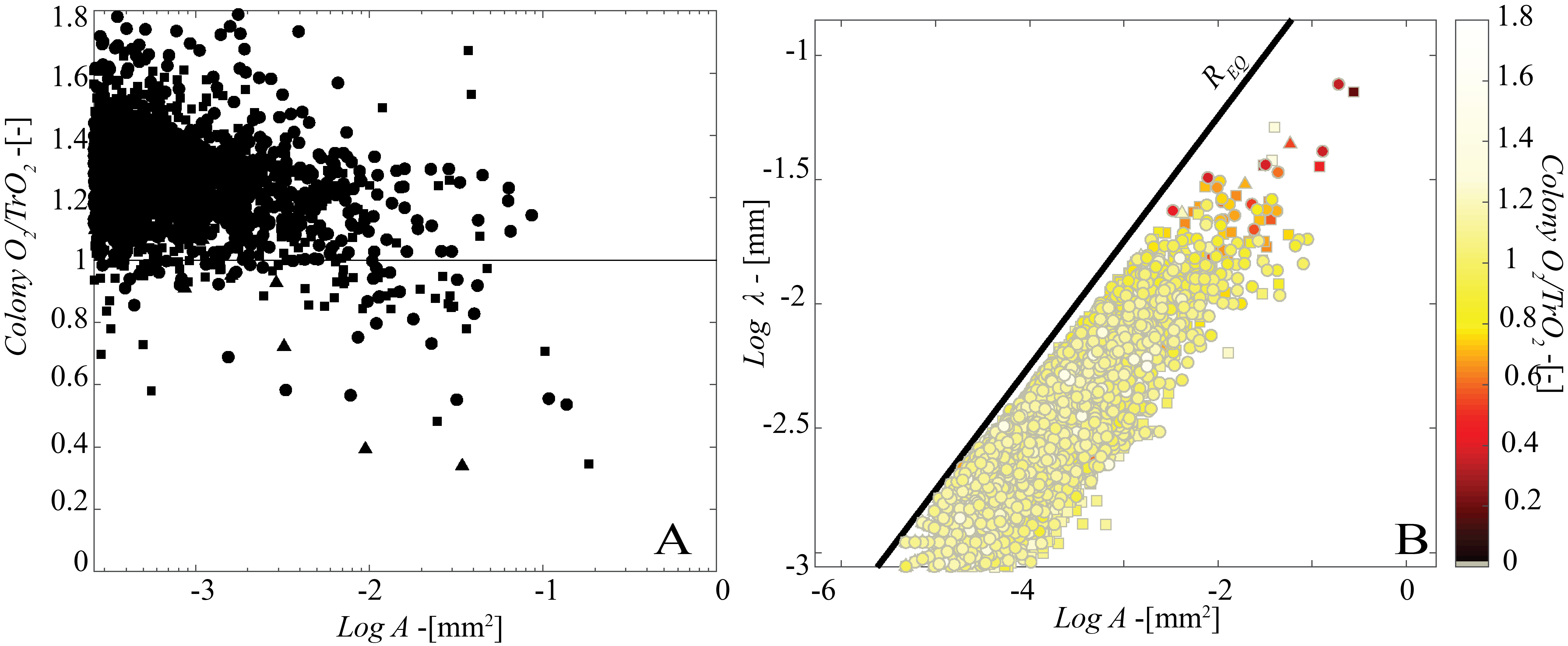}
	\caption{(A) Local O$_2$ concentration observed in each colony ($Colony~O_2$) normalized to the threshold ($Tr_O2$) as a function of the colony area ($A$) at 47~$t_{PV}$ for the O$_2$ impermeable system. (B) Characteristic colony length scale $\lambda$ as a function of the colony area for O$_2$ supply limited to inlet flow at 47$t_{PV}$. Here, marker fill color indicates the local O$_2$ concentration observed in each colony ($Colony~O_2$) normalized to the threshold ($Tr_O2$). In both panels, circles, squares and triangles refer different experiment replica.}
	\label{fig4}
\end{figure}

\noindent
We expect that the colony area (or volume in 3D~systems) is a key factor controlling oxygen uptake: larger areas imply the presence of more bacteria and, likely, faster nutrient an O$_2$ consumption. However, comparing local O$_2$ concentration within each colony (labeled $Colony~O_2$, normalized to the threshold $TrO_2$) to colony area $A$ in Figure~\ref{fig4}A, we found that colony area~$A$ cannot fully explain the formation of anoxic micro-environments which are observed in colonies of any size. Colony oxygen value appears instead to be controlled by the combination of colony planar area and the characteristic length $\lambda$ (or colony volume-to-surface ration in 3D systems), related to colony morphology (Figure~\ref{fig4}B). Fixed the area, colonies with low O$_2$ concentration are characterized by higher $\lambda$ which corresponds to almost rounded shape (or spherical shape in 3D).\\

\noindent
\textbf{Morphology impacts O$_2$ diffusion into bacterial colonies}.\\
The importance of colony morphology in many ecological processes, such as antibiotic resistance and genetic diversity, has been widely well-recognized~\cite{melaugh2016shaping,martinez2022roughening}. This has encouraged an important research effort to understand key factors controlling the dynamics of colony expansion and the  consequent morphology. Among these, the spatial and temporal distribution of nutrients is pivotal: the combined effect of limited nutrient availability in the bulk system, the diffusion process and bacterial nutrient uptake guide the colony morpho-dynamics~\cite{young2022pinning,martinez2022roughening}. As a feedback, the growing pattern modifies physical and chemical properties, such as O$_2$ concentrations, in the colony core and its surrounding environment. This previous research, involving numerical simulations and laboratory experiments, has nonetheless focused on stagnant systems~\cite{farrell2013mechanically,young2022pinning,melaugh2016shaping}. Here, we propose that the role played by colony morphology remains crucial also under flow conditions in complex porous geometries. Our data set, with the simultaneous observation of O$_2$ and biomass maps, allows to directly assess the impact of morphology on anoxic micro-environments formation. This has been done by investigating the two mechanisms regulating O$_2$ dynamics at the pore scale: the O$_2$ uptake associated with aerobic metabolism and the diffusion process transporting O$_2$ from the pore water to the colony core. \\

\begin{figure}
	\centering
	\includegraphics[width=\textwidth]{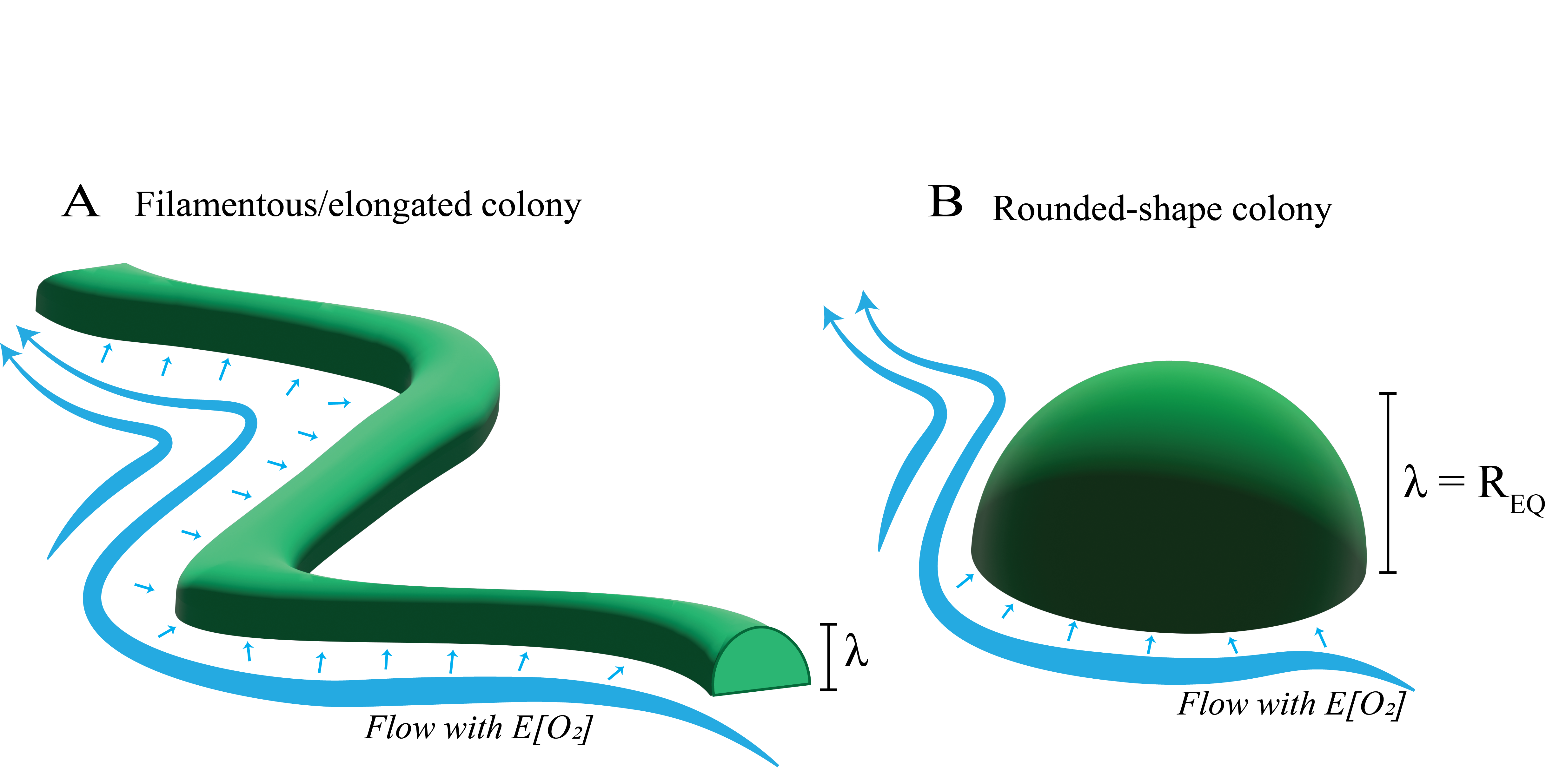}
	\caption{Schematic representation of two paradigmatic colony morphology observed during the experiment: filamentous colony simplified as a hemicylinder (A) and rounded shape simplified as a hemisphere (B).}
	\label{fig5}
\end{figure}

\noindent
To this end, we simplified the colony morphology into an elongated hemicylinder (for filamentous colonies, Figure~\ref{fig5}A) or a hemisphere (for rounded ones, Figure~\ref{fig5}B). These colonies lay on sensor surface and the outer surfaces are in contact with flowing O$_2$-rich pore water. The characteristic length $\lambda$ represents the shortest distance between the colony core and the surrounding pore water which corresponds to the diffusive path for nutrients and oxygen within the colony.\\

\noindent
It is well-established that bacterial growth takes place preferentially at the outer surface of each colony in contact with the pore water that is nutrients and O$_2$-rich~\cite{young2022pinning}. Even though the thickness of the active layer varies with the bacterial specific growth rate and the bulk nutrient concentration~\cite{young2022pinning}, we can assume that O$_2$ uptake by a colony is first and foremost dependent on its surface area, approximated by planar area A in our 2D images. According to the first Fick's law, the O$_2$ diffusion rate is enhanced by a shorter diffusive path (as this generates steeper gradients). Moreover, the overall O$_2$ diffusive mass flux is favored by a higher colony outer surface which is proportional to the mass transfer among pore water and colony. Both colony surface and diffusive path length depends on colony morphology (see schematic Figure~\ref{fig5}). Given a fixed colony surface, O$_2$ diffusion into rounded colonies is slower than into elongated due to the smaller volume-to-surface ratio (longer diffusion path $\lambda$).\\ 

\noindent
\textbf{Morphology significantly alters O$_2$ mass balance in the colony}.\\
We argue that colony shape alters O$_2$ diffusive flux enough to tip the balance from transport to bacterial O$_2$ uptake triggering anoxic micro-environments formation. We therefore propose a quantitative relationship between O$_2$ biomass consumption rate ($R_C$,~[mol~O$_2$/s]) and O$_2$ diffusion rate ($R_T$,~[mol~O$_2$/s]. This \textit{ad hoc} Damköhler number (typically defined for abiotic reactions~\cite{deanna2014mixing}) accounts for colony morphology:
\begin{equation}\label{eqn:Da}
	Da (\lambda) =\frac{R_C}{R{_T}}= \frac{\mu_B \, Y\, \rho}{D_m \, \nabla c_{O_2}(\lambda) A}.
\end{equation} 
\noindent
The external surface area of the colony that should appear in $R_T$ according to the first Fick's law is approximated in Eq.~\ref{eqn:Da} by the planar area $A$ observable from biomass maps.\\

\noindent
The quantity $\mu_B$~[mm$^3$/s] measures the colony growth rate as volumetric change. Under the assumption that colony growth takes place mainly on the outer surface~\cite{vulin2014growing}, $\mu_B$ is proportional to the colony area increasing rate ($\delta~A/\delta t$) and the to volume-to-surface ratio of an individual bacterium during exponential phase ($\sim 1\cdot10^{-3}$~mm, assuming a cylindrical cell with a basal radius of $\sim$~0.6~$\mu$m). Then, $\mu_B$ of each colony is estimated from temporal behavior of colony area observed in biomass maps.\\

\noindent
The O$_2$ concentration gradient between the pore water and colony core, $\nabla c_{O_2}(\lambda)$ [mol~O$_2$/mm\textsuperscript{4}] is expressed as an explicit function of colony morphology through~$\lambda$. It scales as the ratio $ E[O_2]/\lambda$, being $E[O_2]$ the O$_2$ concentration in pore water (assumed equal to the bulk one) and $\lambda$ the diffusion path length determined by colony morphology (see Figure~\ref{fig5}). The three parameters $\rho$~[molC/mm$^3]$ , $Y$~[molO$_2$/molC], and $D_m$~[mm$^2$/s] quantify, respectively, the amount of carbon per volumetric unit of biomass, the bacterial O$_2$ demand and O$_2$ molecular diffusion coefficient as they appear in the literature ($\rho$~=~4.6~10\textsuperscript{-5}~molC/mm$^3$ for \textit{P. putida} strains; $Y$~=~11~molO$_2$/molC; $2\cdot10^{-3}$~mm$^2$/s)~\cite{fagerbakke1996content,zakem2017theoretical}.\\


\begin{figure}
	\centering
	\includegraphics[width=\textwidth]{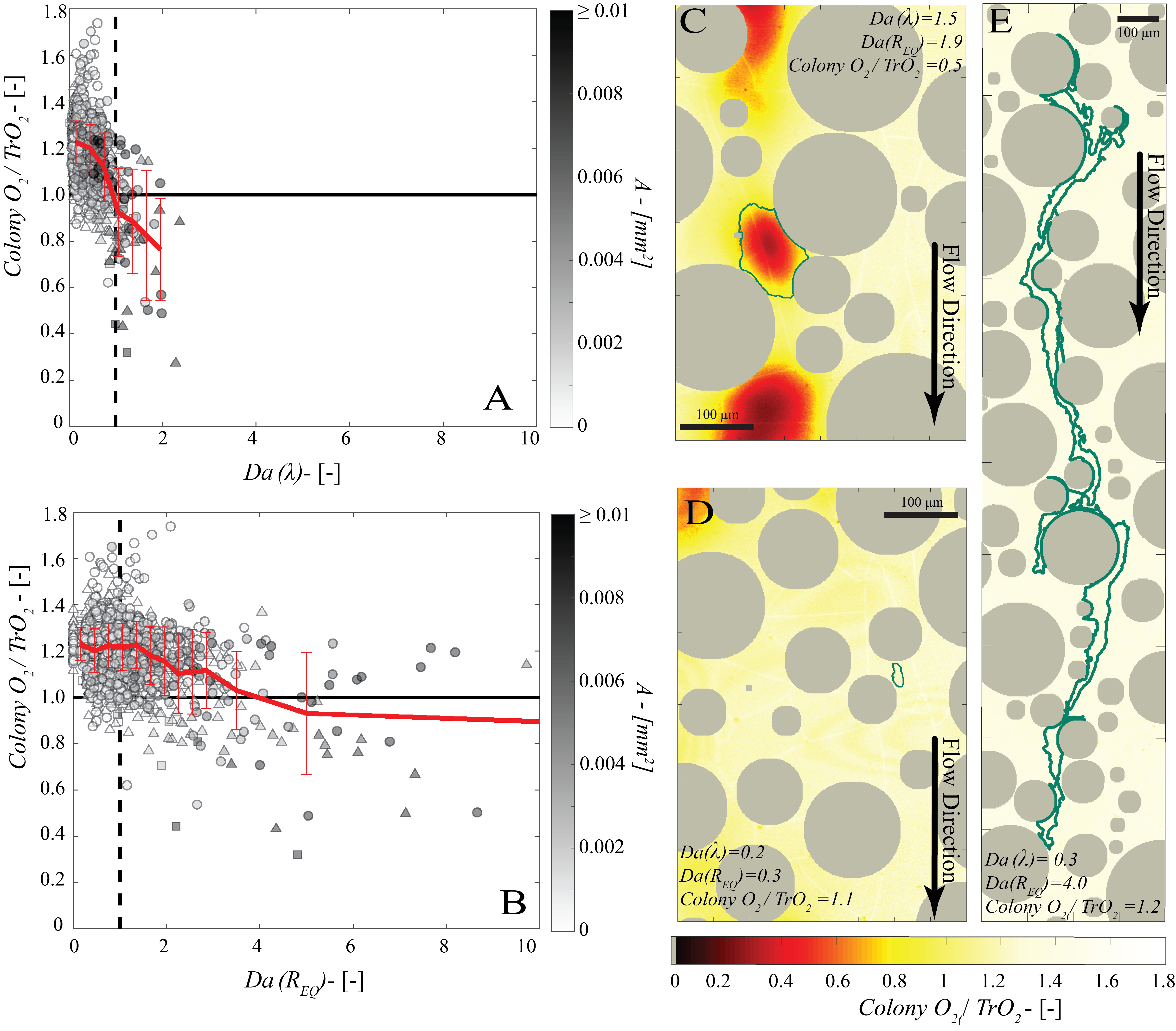}
	\caption{(A) The observed local O$_2$ concentration within each colony (normalized by the threshold $Colony~O_2/Tr_{O_2}$) against the computed value of $Da(\lambda)$. Data presented are averaged between 25 and 65~$t_{PV}$ during which micro-environments persist. Time variability is discussed in SI, section S7. The dashed vertical line denotes the $Da = 1$ which distinguishes O$_2$ diffusion dominated colonies ($Da~<~1$) from conditions suitable for anoxic micro-environments formation ($Da~>~1$). The horizontal solid lines represents the threshold ($Colony~O_2/Tr_{O_2}$) used to discriminate between well-oxygenated and anoxic colonies. Marker face colors indicate the colony area~$A$. The solid red line highlights the average trend of data and the associate standard deviation (vertical errorbars) obtaining by bining $Da$ data into sequential intervals and computing the mean and the standard deviation of $Colony~O_2/Tr_{O_2}$ data within each bin. (B) Same as panel A with $Da$ computed by replacing $\lambda$ with $R_{EQ}$ in Eq.~\ref{eqn:Da} and indicated as $Da(R_{EQ})$. 
	(C-E) Zooms of the O$_2$ spatial map (normalized to the O$_2$ threshold $Tr_{O_2}$) at 47$t_{PV}$ of two rounded colonies of different sizes (C and D) and an elongated colony (E). The colony location is identified by its perimeter (green line). Value of $Da(\lambda)$, $Da(R_{EQ})$ and local oxygen concentration are reported.}
	\label{fig6}
\end{figure}

\noindent
Our results suggest that morphology quantitatively impacts O$_2$ balance inside bacterial colonies since $Da(\lambda)$ spans from 0.4 to 2.5 for colonies of similar areas (Figure~\ref{fig6}A). Local O$_2$ concentrations observed in colonies is well-explained by the conceptual model proposed where $Colony~O_2$ decreases, on average, for increasing $Da(\lambda)$, as clearly highlighted by the average data trend (solid red line in Figure~\ref{fig6}A). The latter shows a marked decreasing behavior of $Colony~O_2$ for increasing $Da(\lambda)$ and falls below the O$_2$ threshold when $Da(\lambda)>1$. According to our metric, conditions compatible with anoxic micro-environments formation (i.e., $Da(\lambda)>1$) are expected in the 1\% of the pore space which is consistent with the value of pore space occupied by anoxic micro-environments obtained from spatial map analyses ($PS_N$,~Figure~\ref{fig3}D). Our metric $Da(\lambda)$ correctly predicts the observed local O$_2$ concentration in the 96\% of the colonies analyzed (see e.g., Figures~\ref{fig6}C-D). A few colonies present morphological features leading to $Da(\lambda)~>~1$ but maintain relatively high local oxygen levels ( $1~<~Colony~O_{2}/Tr_{O_2}~<~1.2$). This misinterpretation is limited to a minor percentage of colonies~(about 0.6\%) and their local O$_2$ concentration was still lower than the bulk one~$E[O_2]$.\\

\noindent
This is further confirmed by the computed average colony O$_2$ behavior (solid red line in Figure 6$B$) as a function of $Da(R_{EQ})$: colony O$_2$ for $Da(R_{EQ})>1$ attains similar mean values of colonies characterized by $Da(R_{EQ})<1$: in other words, it is much less sensitive to $Da(R_{EQ})$ variations.
Neglecting morphology (i.e. considering $Da(R_{EQ})$) leads to overestimating the occurrence of anoxic micro-environments to 30\% of the colonies. Indeed, $Da(R_{EQ})~>~1$ are associated with extended filamentous colonies (see e.g., Figure~\ref{fig6}E) which remain locally well-oxygenated.
We finally highlight that the robustness of the $Da(\lambda)$ metric has been tested with different $Tr_{O_2}$ and the crucial role played by morphology is invariant to the O$_2$ threshold imposed for the anoxic micro-environments identification (see SI, section 5). \\

\noindent
\textbf{Environmental Implications}
Our results show that the formation and persistence of anoxic micro-environments (up to~2\% of the pore space) occur in saturated heterogeneous porous environments despite the continuous injection of well-oxygenated flow. This evidence suggests that excluding the alternative metabolisms of facultative anaerobic bacteria in saturated porous media based on their bulk oxygen content (still commonly done in geochemical modeling of the subsurface environments) would lead to an oversimplification and underestimation of the facultative metabolism contribution to element cycling and contaminant attenuation/mobilization. Our work, using a simplified scenario, provides new insights on the pivotal role of colony morphology on local physical-chemical properties under flow conditions that might help explaining the tiny portion of the porous space occupy by anoxic micro-environments observed in well-oxygenated real subsurface system. Moreover, the proposed metric $Da(\lambda)$, can be adjusted to describe the strain of interest and combined with dedicated models interpreting bacterial colony spatial organization in porous media~\cite{tang2013improved}. This will constitute a preliminary predictive tool to identify bacterial colonies that are prone to become anoxic micro-environments and estimate the portion of anoxic pore space in well-oxygenated porous systems.\\

\noindent
Though the proposed metric can be, in principle, applied to colonies in complex 3D systems, its validity is yet to be tested in such scenarios. Our experimental set-up is limited to 2D geometries representing the pore size heterogeneity and not other structural features typical of natural sediments, e.g., variable surface roughness, the composition of the grains, and nutrient-rich solid aggregates. Moreover, the microbial ecology considered in this study is highly simplified since it accounts for and investigates a laboratory isolate. Still, the strain used in this work belongs to the naturally widespread gender of Pseudomonas. Thus, its behavior might be similar to other strains found in nature, especially if we consider that, according to recent studies~\cite{martinez2022roughening}, colony morphology might be less sensitive to bacterial strain than expected. A large number of factors that might play a key role in anoxic micro-environments formation are yet to be explored, including the use of a more complex bacterial consortium, different porous geometries, and initial/boundary conditions (e.g., pore space saturation level, alternation of stagnant and flow periods, different nutrient spatial and temporal concentration distribution). \\

\noindent
Moreover, longer experiment runs are needed to investigate the possible intermittent behavior of anoxic micro-environments~\cite{kravchenko2017hotspots} and the effect of flow shear stress on the stability of the bacterial clusters. Our innovative methodology, merging easily manipulable microfluidic devices with transparent planar sensors, is flexible and allows individual exploration of all these factors in a fully-controlled laboratory environment. Finally, we highlight that our methodology may find valuable applications to all those disciplines interested in O$_2$ heterogeneity such as bioremediation engineering~\cite{schramm1999occurrence} and biomedical/veterinary studies~\cite{jensen2017microenvironmental}.

\section{Acknowledgements}
	Pietro de Anna acknowledges the support of Swiss National Science Foundation, Project No. $200021\_ 172587$, \emph{Flows in confined micro-structures: Coupling physical heterogeneity and biochemical processes}. Giulia Ceriotti acknowledges support from the \emph{Swiss FCS Postdoctoral Excellence Fellowship}.

\begin{suppinfo}
Details on O$_2$ planar sensors function and calibration; description of the procedures employed to process images collected during the microfluidics experiments; description of the adjusted Pearson Correlation Coefficient used in this study to characterize spatial correlation of biomass and O$_2$ depletion; high-resolution images of biomass and oxygen maps at different times for all replicates of the coated system; discussion about temporal variability of Damköhler number; high-resolution images of biomass and oxygen maps at different times for all replicates of the uncoated system and analyses of results associated with this control scenario. 
\end{suppinfo}

\bibliography{EST-bib}

\end{document}


\section{S1: Preparation of the planar sensor}
An amount of 4 mg of a lipophilic coumarin dye (Bu3Coum)\textsuperscript{[1]}1, 2 mg of platinum(II) meso-pentafluorophenyl porphyrin (Frontier Scientific) and 400 mg of polystyrene (MW 250,000 Da, Acros Organics) were dissolved in 1.6 g of anisole. The obtained solution was screen-printed onto a glass slide with a custom made mask ordered from Welle Oberkirch GmbH (Oberkirch-Zusenhofen, Germany). The glass slides were modified with chlorotrimethylsilane (Aldrich) prior to use.

\section {S2: Optode functioning and calibration}

The optode used in this work is a homogeneous solution of two luminescent dyes in solid matrix polymer (polystyrene). One of the dyes (fluorescent coumarin dye) is not sensitive to oxygen and it constitutes a reference signal. The emission of the second dye (a phosphorescent Pt(II) porphyrin), instead, is quenched by molecular oxygen. The degree of quenching is proportional to the concentration of molecular oxygen. \\

\begin{figure}
	\centering
	\includegraphics[scale=0.8]{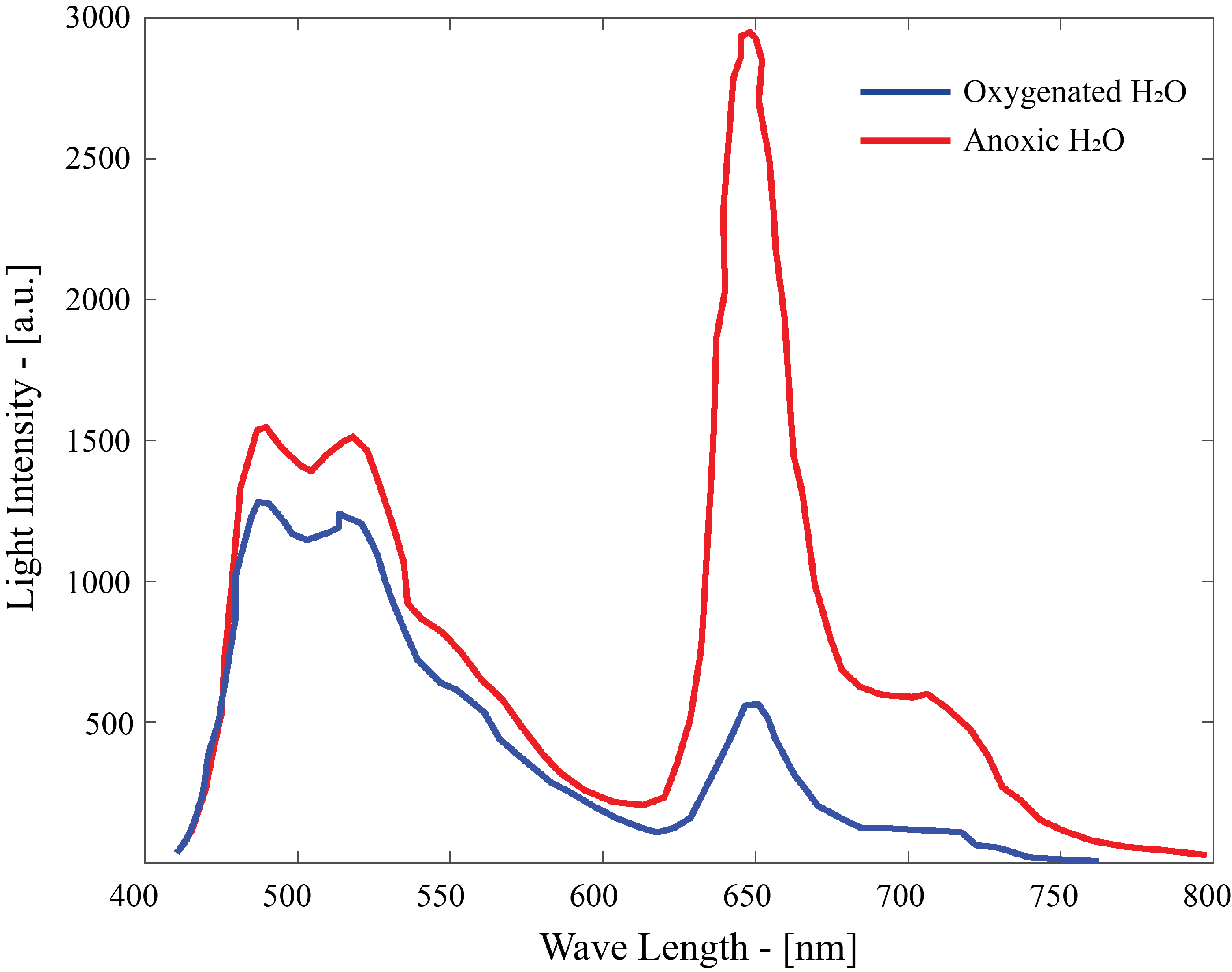}
	\caption{Emission spectra of the optodes in an air-saturated solution (blue line) and anoxic solution (red line).
	}
	\label{figS1}
\end{figure}

\noindent
Both dyes are excited by the same wave length of $450$~nm while emitting their luminescence at two different wavelengths. As such, the overall optode emission signal is a superimposition of two spectra and presents two clearly distinguishable fluorescent peaks. Figure S.\ref{figS1} shows the emission spectrum of the optode for an air-saturated and a anoxic water solution. As we can see, the fluorescent peak located at $500$~nm shows a slight variation that is associated to external ambient factors and not to the oxygen availability, while the peak located $660$~nm remarkably drops when the anoxic solution is replaced by the air-saturated one. \\

\noindent
The O$_2$ concentration assessment is estimated by computing the ratio $I_R$ between the intensity $I_{O2}$ of the $660$~nm peak and the intensity $I_{ref}$ of the $500$~nm reference peak. In this way, eventual emission fluctuations associated with factors other than O$_2$ concentration variation are removed since they would affect in the same way both signals $I_{O2}$ and $I_{ref}$.
\\

\begin{figure}
	\centering
	\includegraphics[scale=0.8]{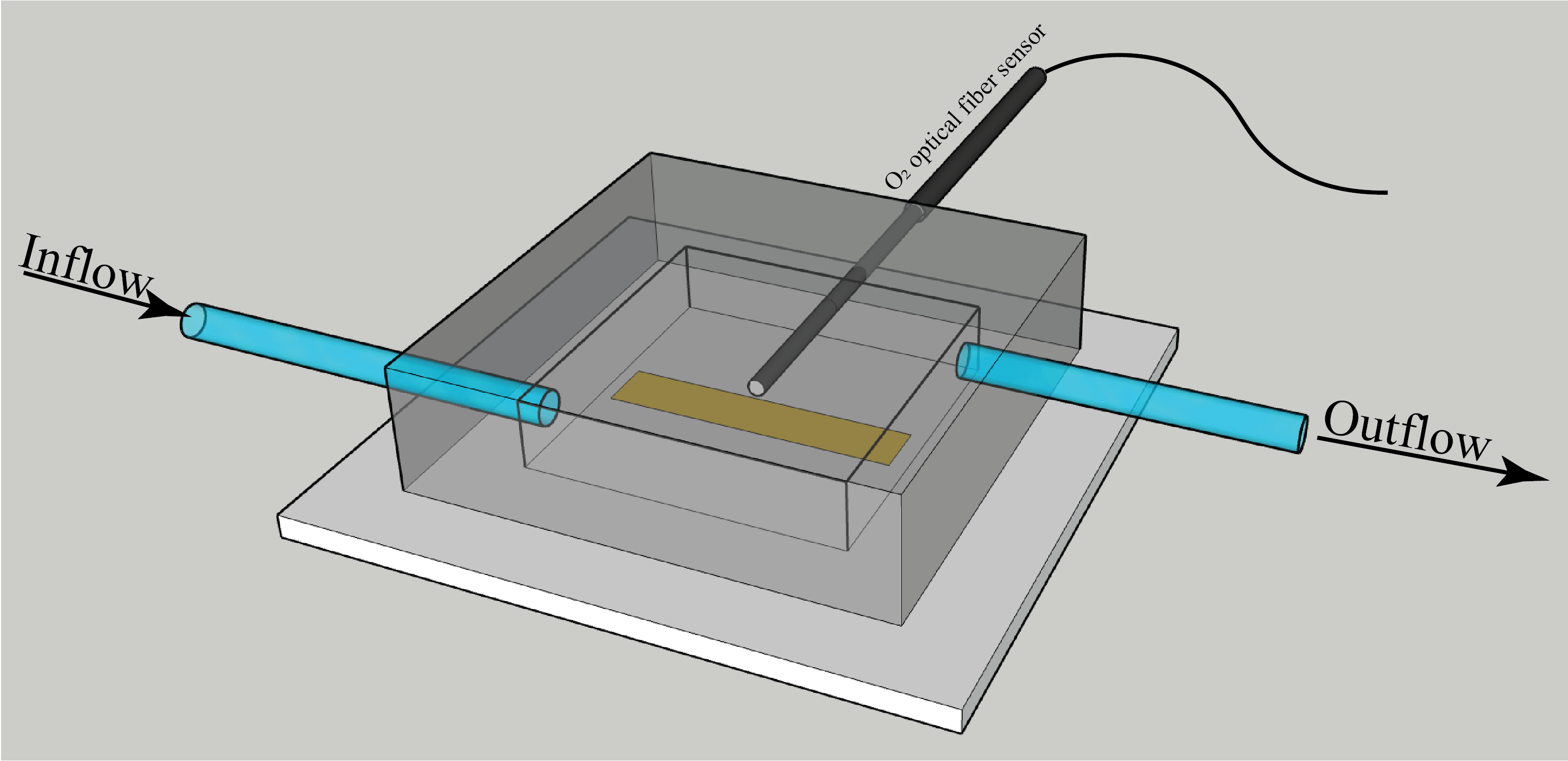}
	\caption{Outline of the chip used for optode calibration procedure.}
	\label{figS2}
\end{figure}

\begin{figure}
	\centering
	\includegraphics[scale=0.8]{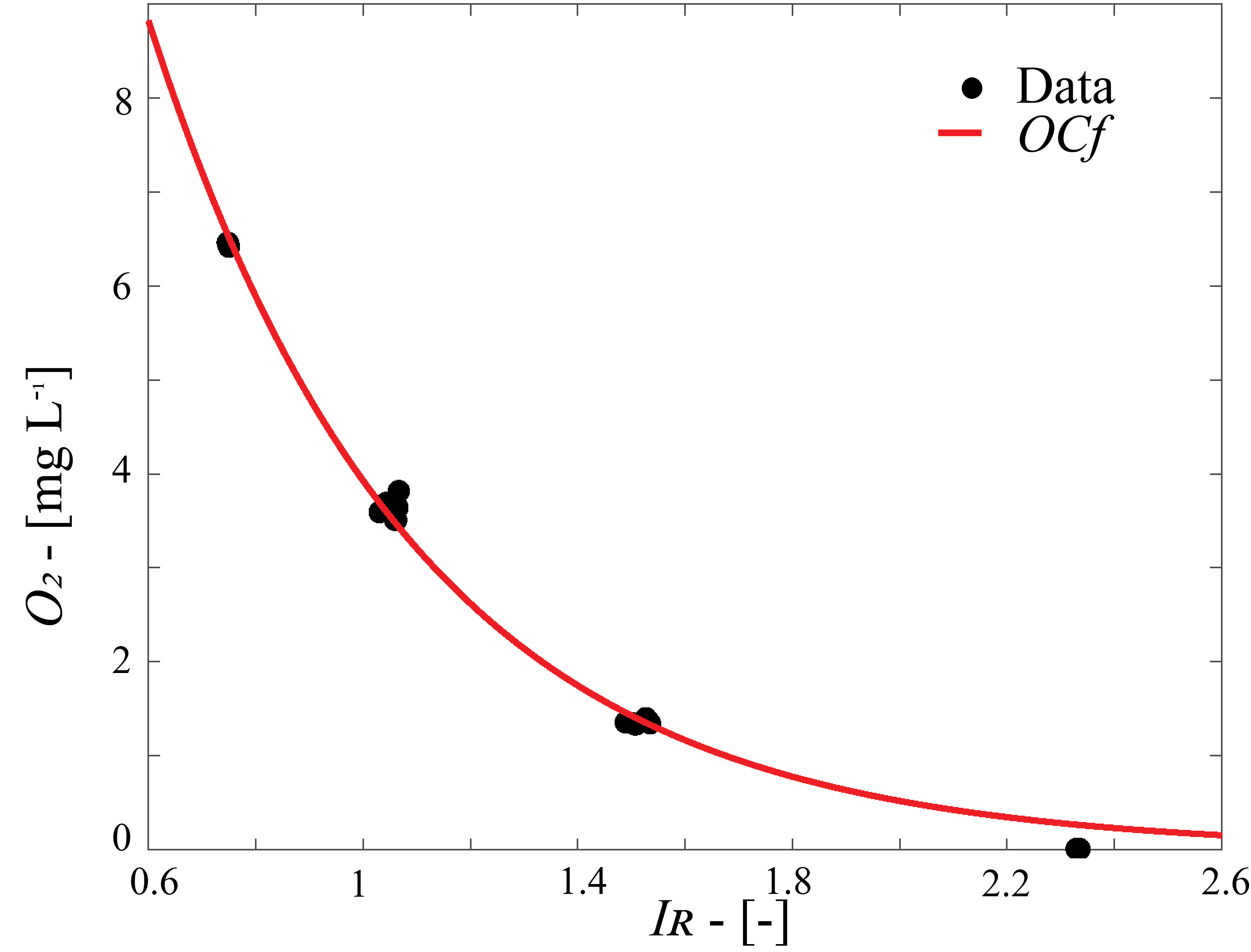}
	\caption{Calibration data (black circles) collected during the optode calibration procedure and the fitting curve.}
	\label{figS3}
\end{figure}

\noindent
As shown in Figure~S\ref{figS2} chip (length $55$~mm, width $15$~mm and thickness $3.5$~mm), modeled by purring polydimethylsiloxane (PDMS; Sylgard 184 Silicone Elastomer Kit, Dow Corning, Midland, MI) with the addition of 10 w/w \% of curing agent in a plastic mould, is plasma-bonded on top of a microscope glass slide where an optode has been deposited via screen-printing technique. Two Tygon tubes are connected to the chip allowing injecting a chosen solution in the chip chamber. The entire chip surface has been coated with a layer of NOA-81 glue to prevent oxygen in-fluxing from the external atmosphere. A fiber optic O$_2$ sensor protected by a sharp syringe needle (TROX430, Pyroscience) is speared into the PDMS layer constituting the wall of the chip chamber such that the measuring point is located in close proximity to the planar optode sensor as outlined in Figure~S.\ref{figS3}. Before the chip set-up, the fiber optic O$_2$ sensor is calibrated according to guidelines provided by the supplier. \\

\noindent
Four water solutions with different dissolved O$_2$ concentrations are prepared by mixing air-saturated Milli-Q water ($S_{high}$, obtained by carefully shaking three times a 250 mL Schott bottle filled with 100 mL of Milli- Q water and opening the cap after each shake for few minutes) and a poorly oxygenated Milli-Q water ($S_{low}$) obtained by dissolving $\simeq$ 8 mg of sodium sulfate in 100 mL of Milli-Q water leading to a solution with O$_2$ concentration equal to 0.1 mg/L. The chemical O$_2$ consumption by sodium sulfate is catalyzed by adding 40 $\mu$L of standard Cobalt solution (Merck, Co(NO\textsubscript{3})\textsubscript{2} in HNO\textsubscript{3} 0.5 mol/l 1000 mg/l Co Certipur\textsuperscript{\textregistered}) before mixing. A fully anoxic solution ($S_0$) is also prepared by adding an excess of sodium sulfate in 100 mL of Milli-Q water. \\

\noindent
The chip is placed on a miscrocope stage, 10 mL of each solution, with different dissolved O$_2$, are separately injected manually into the chip thorough a Benton glass syringe. Two pictures of the optode signal are taken where the optical fiber has been positioned. The first picture captures the reference dye signal by shining the optode with a blue LED ($440 \pm 20$~nm) generated by the Lumencor SPECTRA X Light Engine and filtering the emission signal through a Semrock GFP bandpass emission filter. The second image captures the phosphorescent signal of the O$_2$ sensitive dye obtained by illuminating the chip with a blue LED ($440 \pm 20$~ nm) generated by the same LED and filtering the emission signal through a custom Semrock single-band band pass filter $662 \pm 11$~nm. Four
couples of pictures are taken to test the stability of our optode measurement. \\

\noindent
Figure~S.\ref{figS3} shows all data collected: the dissolved O$_2$ concentration measured by the TROX430 are reported against the signal ratio $R_I$ detected for each couple fo pictures. The best fit of these data (solid red line in Figure~S.\ref{figS3}) is the exponential function 
\begin{equation}
	c = 29.7 \, e^{-2.02 \, I_R}.
\end{equation}  
which we found using the MATLAB Curve Fitting App with a confidence level of 95 \% on parameter estimated value. The resulting $R^2$ metric is equal to 0.994. \\

\noindent
Overall, the chosen optode sensor presents three major advantages. First, the presence of the sensor on the glass slide does not alter the chip transparency. Second, the employment of O$_2$ sensitive dye in combination with a reference dye provides a robust measurement of dissolved O$_2$ concentration, which is independent from external environmental conditions and experimental set up (e.g. camera, light source a and objective features). Third, the eventual quenching mechanism does not consume O$_2$.

\section{S3: Image processing}

\textbf{The mask}. We define a binary matrix $M$, of the same size as the images we acquired to store the information on the porous medium structure. To this end, we use the image acquired with an adjusted phase contrast configuration just before microbial suspension inoculation. We used a large phase ring (Nikon Ph3) compared to the ring (a Ph1) present in the objective, so that the sample gets illuminated almost laterally. This results in a dark background and the perimeter of each grain will appear as a bright circle. First, we set to $0$ the value of each pixel below a local threshold value defined by the Matlab function \emph{adaptthresh}: it corresponds to all pixels other than grains' perimeter. Second, exploiting the Matlab function \emph{imfill}, we assign to all pixels inside a circle a value of $1$, obtaining disks of the same size and location as the physical grains. The mask $M$ is then derived as the inverted image of the latter matrix:  pixels representing solid grains are associated with the value of $0$, and $1$ otherwise. \\

\noindent
\textbf{Microbial spatial distribution}. Images for biomass distribution were taken with an adjusted phase contrast optical configuration used for mask. In this configuration, bacteria appear as bright spots. To quantify the distribution of bacteria $B_t$ at each acquisition time $t$, we consider the matrix $I_t(x,y)$ corresponding to the image acquired with the adjusted phase contrast configuration, where $x$ and $y$ represent the longitudinal and transverse spatial directions. First, each element of $I$ is rescaled between 0 (dark pixels) and 1 (white pixel) dividing its value by $2^{16}-1$ (i.e. the camera pixel depth). Second, we subtract to the rescaled $I_t(x,y)$ the matrix $I_0(x,y)$ acquired just before bacteria inoculation (at time $t=0$). We set to $0$ all negative pixels of the resulting image. Third, we multiply, element by element, this matrix by the defined mask, i.e., $B_t = I(x,y) \, M(x,y)$, obtaining the local biomass distribution.\\

\noindent
\textbf{Pore space occupied by biomass ($PS_B$) and anoxic microenvironments ($PS_N$)}.
The quantification of the pore space occupied by biomass is performed starting from the maps $B_t$. Each biomass map was transformed into a binary map $BinB_t$ by assigning 1 to all pixels with a value of $B_t > 0$. This means that we consider all pixels where light is scattered compared to the image at $t=0$ as occupied by newly grown biomass. The value of $PS_B$ is then computed as 

\begin{equation}
	PS_B = \frac{\Sigma{BinB_t}}{\Sigma{M}}100
\end{equation} 

Similarly, to identify anoxic microenvironments, the O$_2$ maps ($c_{O_2}$) were transformed into binary images $Binc_{O_2}$ by associating with the value of 1 all pixels where $c_{O_2} < TrO_2$ and 0 otherwise.
The value of $PS_N$ results from
\begin{equation}
	PS_B = \frac{\Sigma{Binc_{O_2}}}{\Sigma{M}}100
\end{equation} 

\section{S4: Details on the adjusted Pearson Correlation Coefficient}
\noindent
To quantitatively correlate the emergence of the microenvironments to the location of bacterial colonies, we compute an adjusted Pearson Correlation Coefficient (PCC), defined as
\begin{equation}\label{eqn:Pearson}
	PCC (t) = \frac{\sum_{xy} \big( (O_2^C (x,y) - E[O_2^C] \big) \, \big( B_t(x,y) - E[B_t](t)) \big) }{ \sqrt{\sum_{xy} \big(O_2^C (x,y) - E[O_2^C ] \big)^2 \, \sum_{xy} \big(B_t(x,y) - E[B_t](t) \big) ^2} }
\end{equation}
where the volume averages of biomass $E[B_t]$ was defined above and the complementary oxygen concentration $O_2^C$ is defined as
\begin{equation}
	O_2^C = 1 - \frac{\textrm{max}(c_{O_2}) - c_{O_2}}{\textrm{max}(c_{O_2})}
	\label{eqn:Ocompl}
\end{equation}
and $E[O_2^C]$ represents its overall average. The $O_2^C$ spatial map gets values between $0$ and $1$, and is the complementary spatial map of the O$_2$ concentration being close to $0$ for high oxygen values and close to 1 for low ones. \\

\newpage
\section{S5: Details on $TrO_2$ selection and results robustness}
\begin{figure}
	\centering
	\includegraphics[scale=1]{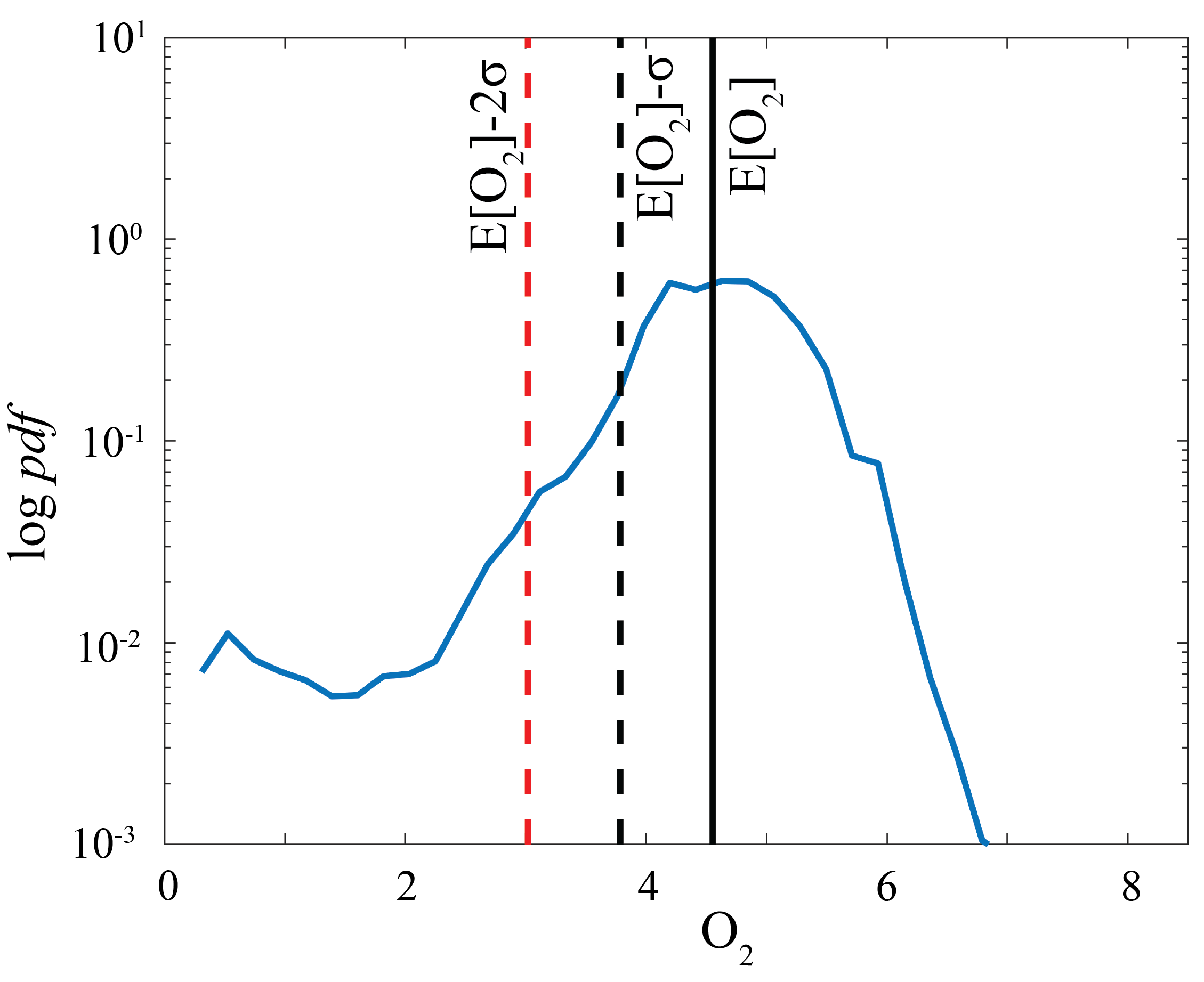}
	\caption{ Probability density function \textit{pdf} of O$_2$ concentrations at 47 tPV in replicate 3 of the experiment. Solid and dashed black lines indicate the average ($E[O_2]$) and the variability around the average value ($E[O_2]-\sigma$) of the \textit{pdf}, respectively. The dashed red line is the chosen threshold for anoxic microenvironment identification.}
	\label{fgr:S3_1}
\end{figure}

A universally shared definition of anoxic microenvironment has not been established yet, as discussed in section "Anoxic microenvironments are correlated to bacterial colony growth in space and time" of the manuscript. The metric $TrO_2$should be adjusted as a function of the strain/process and the environment one is interested in. In this work, we do not explore the activation of any facultative metabolism and, thus, such strain-specific threshold identification cannot be applied and we defined our threshold based on a purely statistical criterion. We have analyzed all oxygen concentration maps in terms of probability density function ($pdf$), average ($E[O_2]$) and standard deviation ($\sigma$). The maximum standard deviation, corresponding to the highest observed variance around the average value, occurs at 47 tPV in replicate 3 of our experiment and settles around 10\% of the average value $E[O_2]$ (see Figure~S\ref{fgr:S3_1}). Acknowledging a component of arbitrariness in the choice, we defined O$_2$ concentrations lower than $E[O_2]$-2$\sigma$ to be significantly different from the bulk value, or, equivalently, set the threshold  $TrO_2$ = 0.8$E[O_2]$. According to the chosen criterion, the anoxic microenvironments constitute 5\textsuperscript{th}-percentile of the $pdf$ (see Figure~S\ref{fgr:S3_1}). This criterion is then transferred to all the O$_2$ maps. 

Still, the results proposed in the manuscript are invariant to chosen $Tr_{O2}$. We have tested the performance of the proposed metric $Da(\lambda)$ (accounting for colony morphology) and $Da(R_{EQ})$ (assuming round colonies) to identify anoxic microenvironment formation. We selected two more stringent $TrO_2$ in addition to 0.8$E[O_2]$, i.e., 0.7$E[O_2]$ and 0.6$E[O_2]$. We assessed the improved predictability of anoxic microenvironment formation led by $Da(\lambda)$ compared to $Da(R_{EQ})$ using as criterion the rate of metric success. We define the latter as the percentage of data for which the observed $Colony$ $O_2/Tr_{O_2}$ is correctly interpreted by the metrics  $Da(\lambda)$ and $Da(R_{EQ})$, i.e., observed $Colony$ $O_2/Tr_{O_2}<$1 associated with $Da>$1 and \textit{vice versa}. The comparison of performances of  $Da(\lambda)$ and $Da(R_{EQ})$ (Table S.\ref{tbl}) clearly indicate that the results presented in the manuscript are insensitive to the chosen $Tr_{O2}$ and accounting for colony morphology provides a better interpretation of the anoxic microenvironment formation for all explored $Tr_{O2}$. 

\begin{table}[]
	\caption{Rate of success and failure of the metrics $Da(\lambda)$ and $Da(R_{EQ})$to predict the formation of anoxic microenvironments for different imposed $Tr_{O2}$.}
	\begin{tabular}{l|ll|ll|}
		$Tr_{O2}$& Rate of success [\%] &  & Rate of failure [\%] &  \\
		&  $Da(\lambda)$& $Da(R_{EQ})$  &$Da(\lambda)$  & $Da(R_{EQ})$ \\
		0.8$E[O_2]$& 96.5& 65.6 & 3.5 & 34.4\\
		0.7$E[O_2]$& 98.5& 66.5 & 1.5 & 33.5\\
		0.6$E[O_2]$& 98.2& 66.6 & 1.8 & 33.4\\
	\end{tabular}
\label{tbl}
\end{table}       

\newpage
\section{S6: Additional results for the experiment with NOA-81 coated chip}

\begin{landscape}	
\begin{figure}
	\centering
	\includegraphics[scale=0.8]{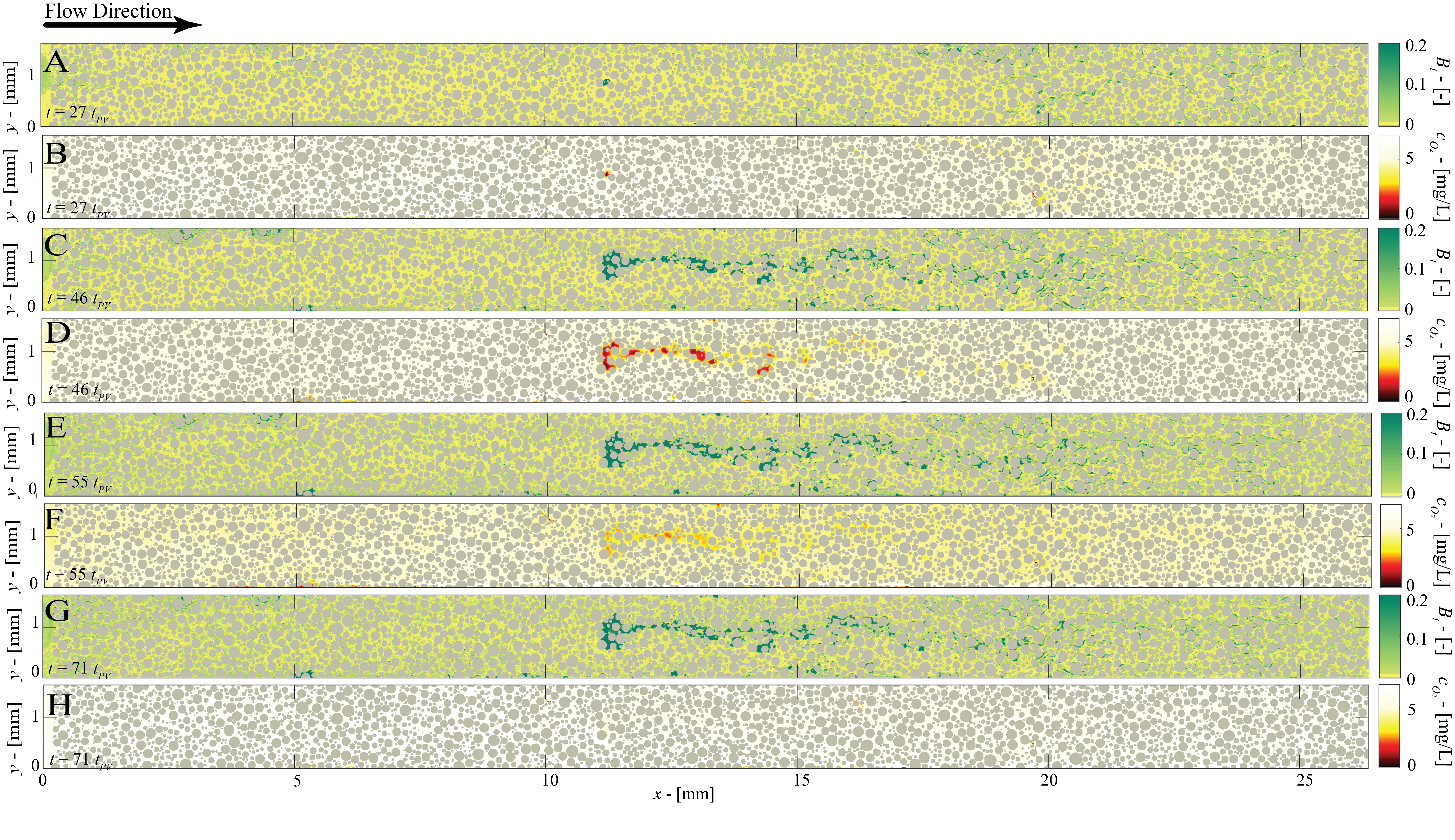}
	\caption{Spatial distribution of $B_t$ for replicate 1 of the experiment with NOA-81 coated chip at $t$ = 27$t_{PV}$ (panel A), $t$ = 46$t_{PV}$ (panel C), $t$ = 55$t_{PV}$ (panel E) and $t$ = 71$t_{PV}$ (panel G) compared to the corresponding $C_{O_2}$ maps ($t$ = 27$t_{PV}$ in panel B, $t$ = 46$t_{PV}$ in panel D, $t$ = 55$t_{PV}$ in panel F, $t$ = 71$t_{PV}$ in panel H).}
	\label{fgr:S10_1}
\end{figure}
\end{landscape}

\newpage
\begin{landscape}
\begin{figure}
	\centering
	\includegraphics[scale=0.8]{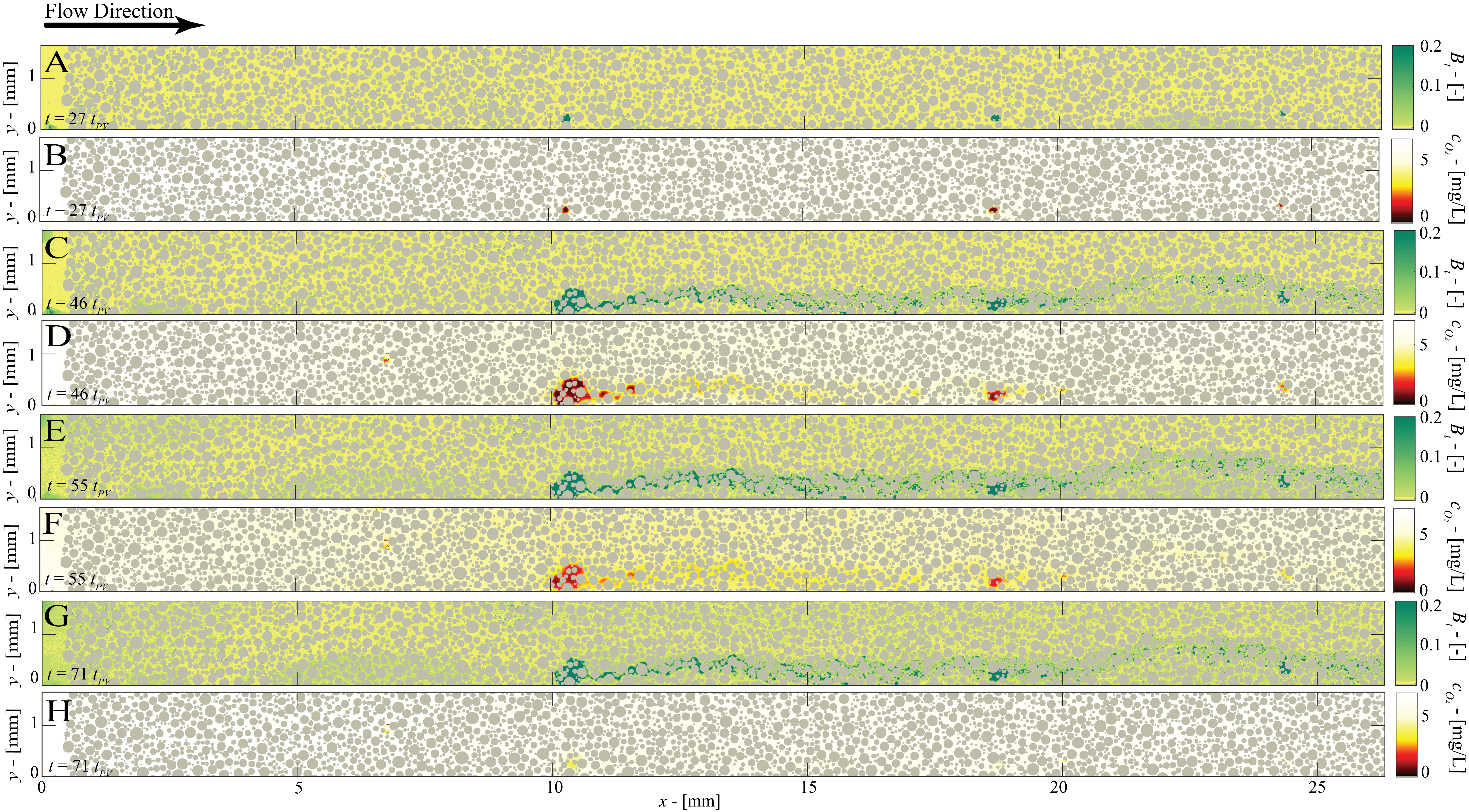}
	\caption{Spatial distribution of $B_t$ for replicate 2 of the experiment with NOA-81 coated chip at $t$ = 27$t_{PV}$ (panel A), $t$ = 46$t_{PV}$ (panel C), $t$ = 55$t_{PV}$ (panel E) and $t$ = 71$t_{PV}$ (panel G) compared to the corresponding $C_{O_2}$ maps ($t$ = 27$t_{PV}$ in panel B, $t$ = 46$t_{PV}$ in panel D, $t$ = 55$t_{PV}$ in panel F, $t$ = 71$t_{PV}$ in panel H).}
	\label{fgr:S10_2}
\end{figure}
\end{landscape}
\newpage
\begin{landscape}
	\begin{figure}
		\centering
		\includegraphics[scale=0.8]{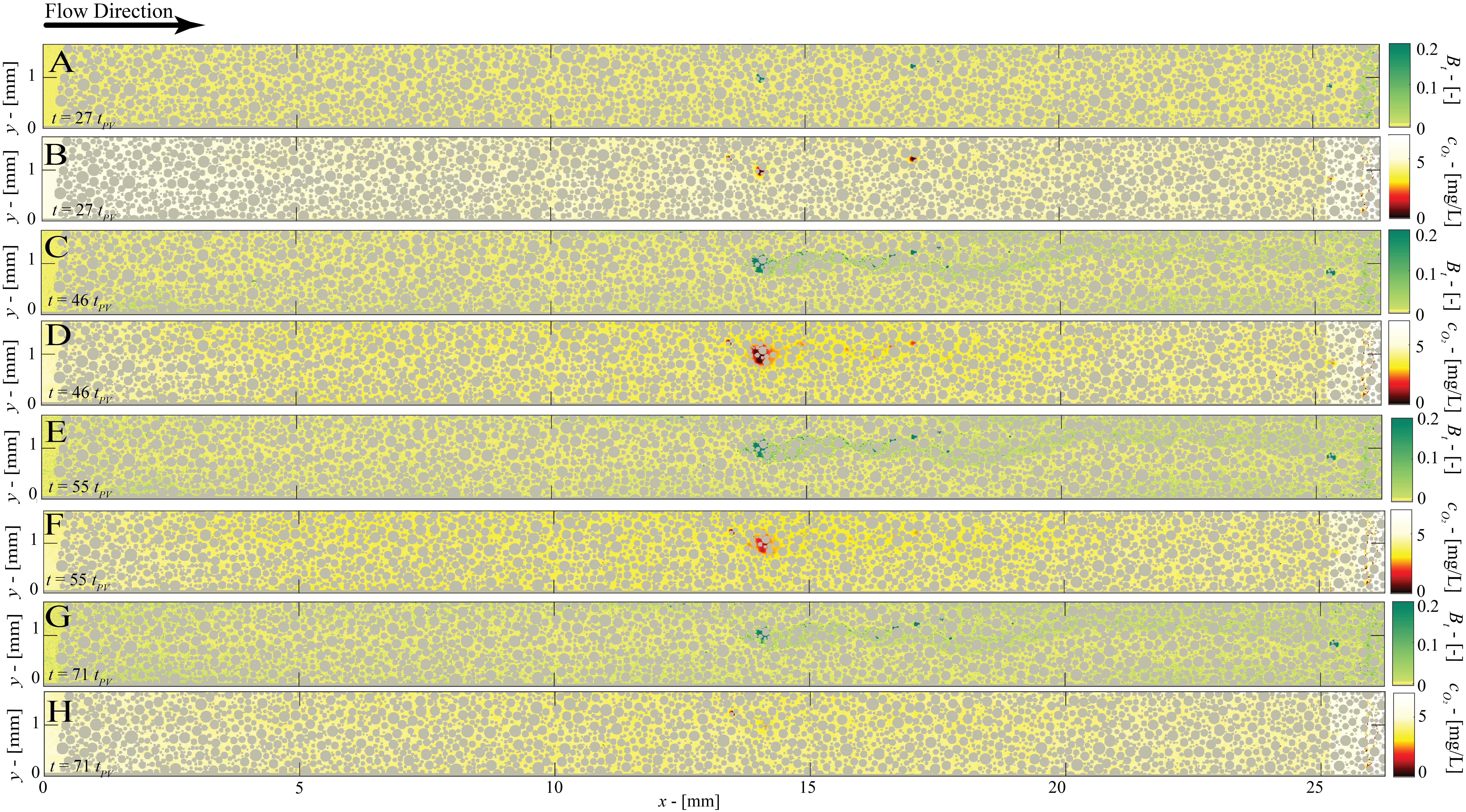}
		\caption{Spatial distribution of $B_t$ for replicate 3 of the experiment with NOA-81 coated chip at $t$ = 27$t_{PV}$ (panel A), $t$ = 46$t_{PV}$ (panel C), $t$ = 55$t_{PV}$ (panel E) and $t$ = 71$t_{PV}$ (panel G) compared to the corresponding $C_{O_2}$ maps ($t$ = 27$t_{PV}$ in panel B, $t$ = 46$t_{PV}$ in panel D, $t$ = 55$t_{PV}$ in panel F, $t$ = 71$t_{PV}$ in panel H).}
		\label{fgr:S10_3}
	\end{figure}
\end{landscape}

\subsection{S7: $Da(\lambda)$ time variability }
\begin{figure}
	\centering
	\includegraphics[scale=0.8]{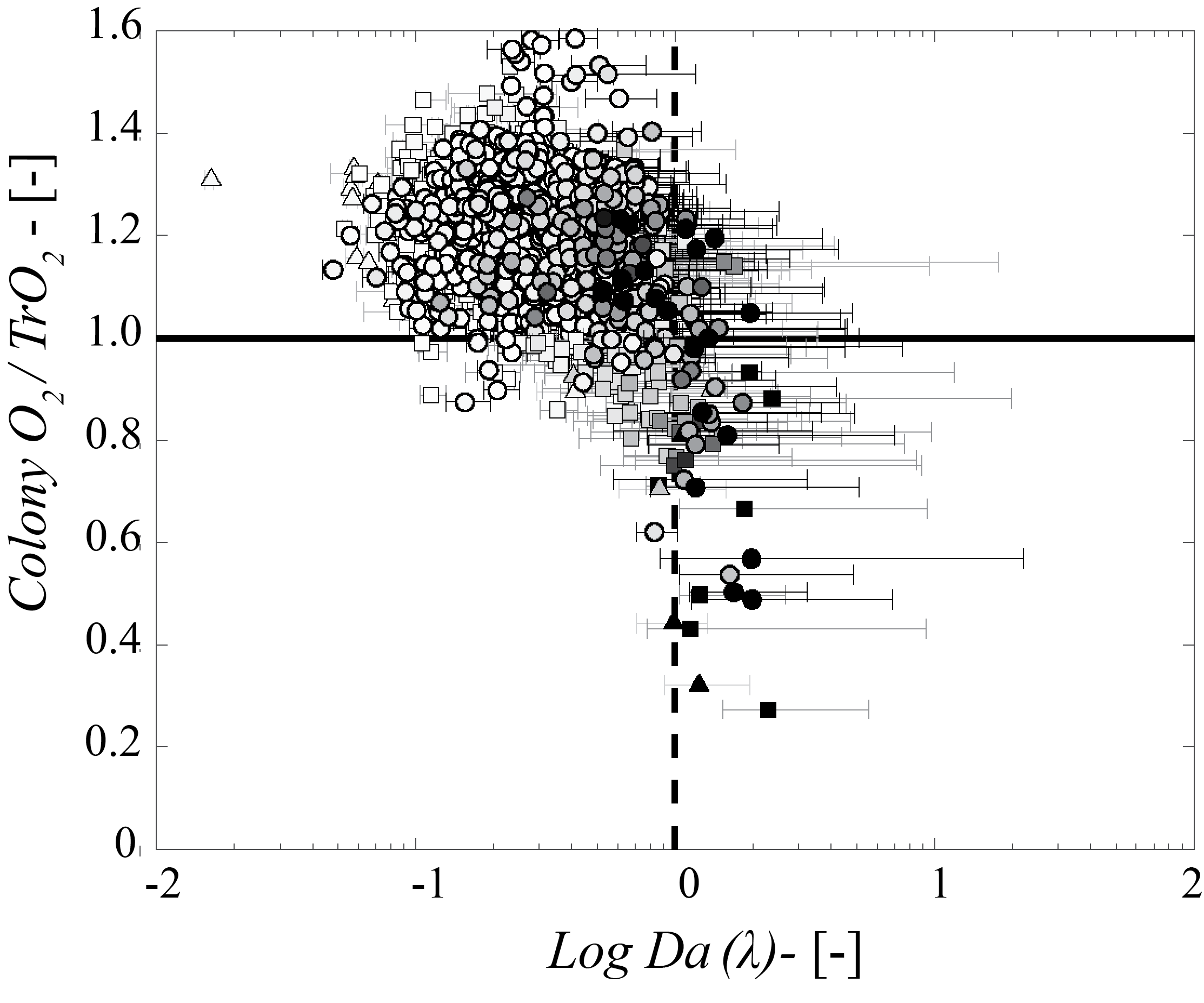}
	\caption{Behavior of $Da$ computed for each biomass cluster using $\lambda$ as O$_2$ diffusion characteristic length scale in as a function of the average value of O$_2$ measured inside each cluster for the control experiment without NOA-81 coated chip. Symbols represent the averaged value of $Da(\lambda)$ over the time interval between 27 and 55~$t_{PV}$, corresponding to what presented in the manuscript. The horizontal bars indicate the range associated with each point, i.e. the minimum and the maximum value observed in the time interval between 27 and 55~$t_{PV}$. The dashed vertical line highlights the $Da = 1$, while the horizontal solid line represents the threshold ($Colony$ $O_2/Tr_{O_2}$) used to discriminate between a well-oxygenated and an anoxic colony. Markers face colors indicate the average $A$ of the colony during the chosen interval of time.}
	\label{Da_time}
\end{figure}

In the manuscript, we present the data of $Da(\lambda)$ as an average between 27 and 55~$t_{PV}$, to capture the mean behavior of the colony in the time interval of interest, i.e., the one associated with anoxic microenvironment formation. We investigated the time variability of the $Da(\lambda)$ as a consequence of $\lambda$ temporal behavior in Figure S.\ref{Da_time} where the mean values presented in the main manuscript are enclosed in their range, i.e. between the maximum and minimum value observed in the time interval for each datum. \\

\noindent
On the one hand, for $Colony$ $O_2/Tr_{O_2}>1.2$, all of the data and the corresponding ranges fall in $Da(\lambda)<1$ indicating that, regardless the observation time, the conditions for the formation of anoxic microenvironments are never met.
On the other hand, colonies that show on average a $Colony$ $O_2/Tr_{O_2}<0.9$ are characterized by a $Da(\lambda)$ that is always greater than one. This leads to the formation of microenvironments with highly depleted O$_2$ concentration in their core. \\

\noindent
Finally, for  $0.9<$ $Colony$ $O_2/Tr_{O_2}<1.2$, we observe a transition zone where the mean $Da(\lambda)$ and the associated range are progressively shifted towards higher values for decreasing level of average O$_2$ concentration. We can deduce that some colonies may have a fluctuating growth behavior in time and, thus, the conditions for the formation of anoxic microenvironments, as defined in our work, are not stably met. As a consequence, these colonies present an average $Colony$ $O_2/Tr_{O_2}$ around one, i.e. close to the chosen threshold. \\

\noindent
To conclude, the overall trend of $Da(\lambda)$ values and their associated temporal ranges is consistent with the observed O$_2$ average concentration, supporting the good performances of $Da(\lambda)$ as a metric for identifying the conditions for anoxic microenvironment formation.  
\newpage
\section{S8: Results for uncoated chip - control scenario}
\begin{figure}
	\centering
	\includegraphics[scale=0.5]{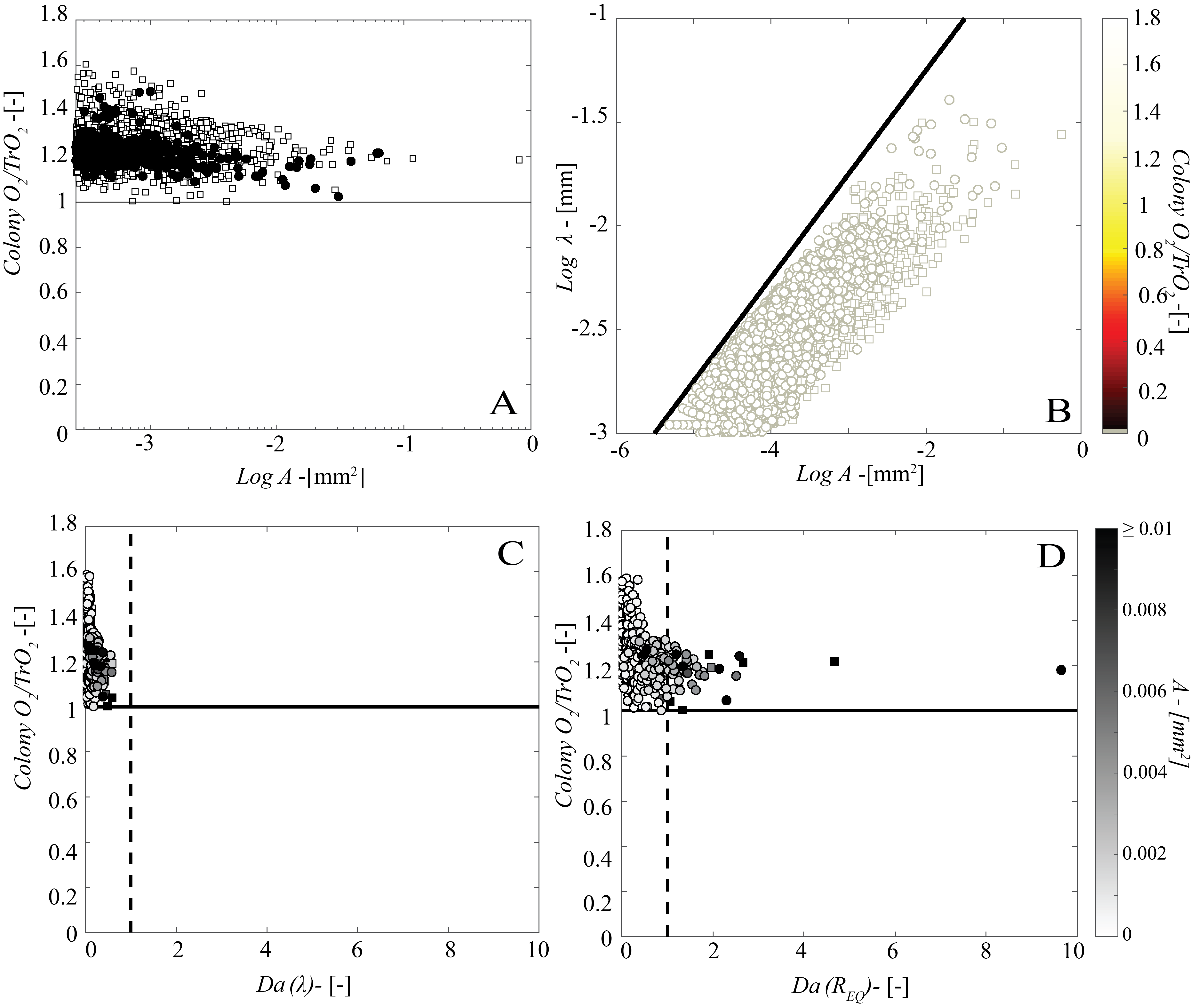}
	\caption{A) Values of the average oxygen level observed in each colony normalized to the threshold used for anoxic microenvironment identification ($Colony$ $O_2/Tr_{O_2}$) as a function of the areas covered by each colony at 47 $t_{PV}$. B) Values of the characteristic colony length scales $\lambda$ as a function of the colony area for the experiment without NOA-81 coating at 47 $t_{PV}$. Markers color indicates the average oxygen level observed in each colony normalized to the threshold used for anoxic microenvironment identification ($Colony$ $O_2/Tr_{O_2}$). In both panels, circles and squares refer to replica 1 and replica 2 of the experiment. In panles C and D, behavior of the average value of $Da$ computed for each biomass cluster using $\lambda$ (panel C) and $R_{EQ}$ (panel D) as O$_2$ diffusion characteristic length scale in $R_T$ of Eq.\ref{eqn:Da} as a function of the average value of O$_2$ measured inside each cluster for the control experiment without NOA-81 coated chip. In both panels C and D, the $Da(\lambda)$ and $Da(R_{EQ})$, respectively, are averaged over the time interval between 27 and 55~$t_{PV}$, corresponding to what presented in the manuscript. The dashed vertical line highlights the $Da = 1$, while the horizontal solid line represents the threshold ($O_{2(A)/Tr_{O_2}}$) used to discriminate between a well-oxygenated and an anoxic colony. Markers face colors indicate the average $A$ of the colony during the chosen interval of time.}
\end{figure}

\newpage

\begin{landscape}
	\begin{figure}
		\centering
		\includegraphics[scale=0.8]{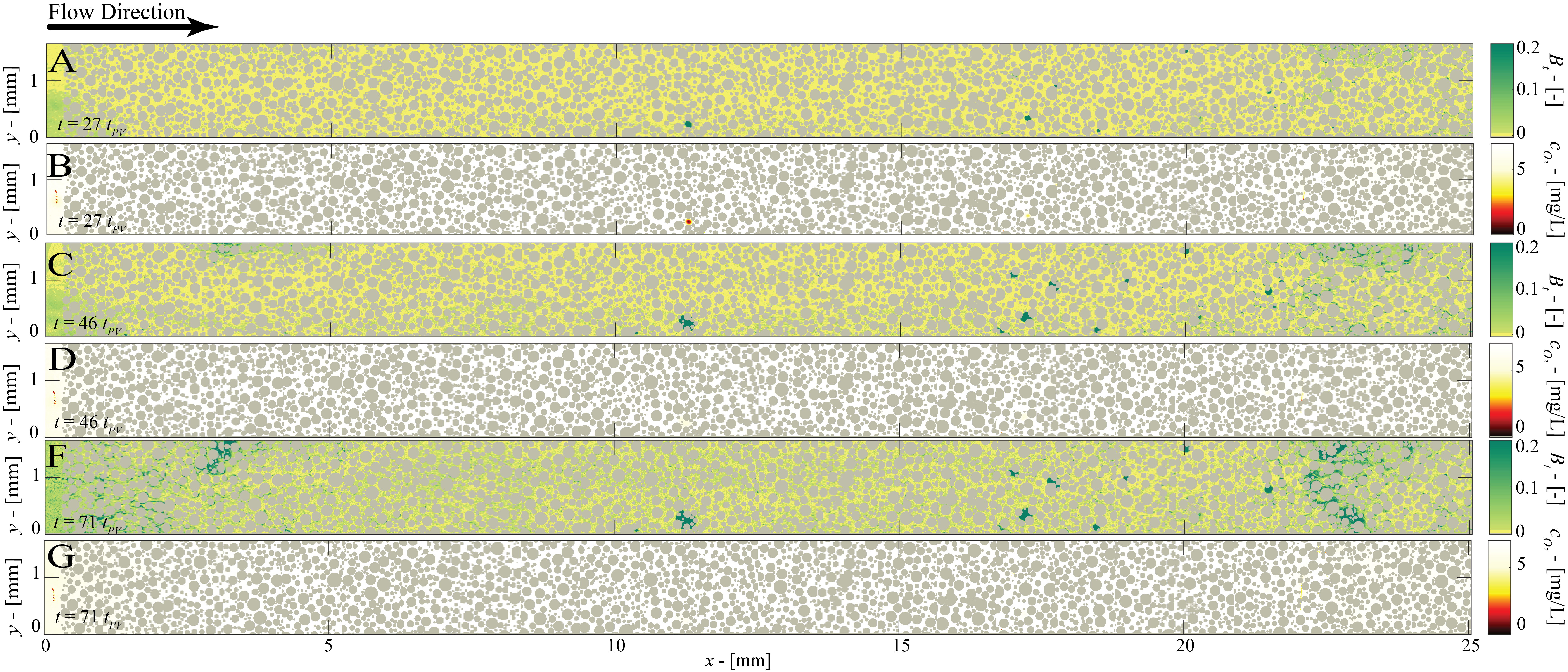}
		\caption{Spatial distribution of $B$ for replicate 1 of the experiment with uncoated chip at $t$ = 27$t_{PV}$ (panel A), $t$ = 46$t_{PV}$ (panel C) and $t$ = 27$t_{PV}$ (panel F) compared to the corresponding $c_{O_2}$ maps ($t$ = 27$t_{PV}$ in panel B, $t$ = 46$t_{PV}$ in panel D, $t$ = 71$t_{PV}$ in panel G).}
		\label{fgr:C10_1}
	\end{figure}
\end{landscape}

\newpage
\begin{landscape}
	\begin{figure}
		\centering
		\includegraphics[scale=0.8]{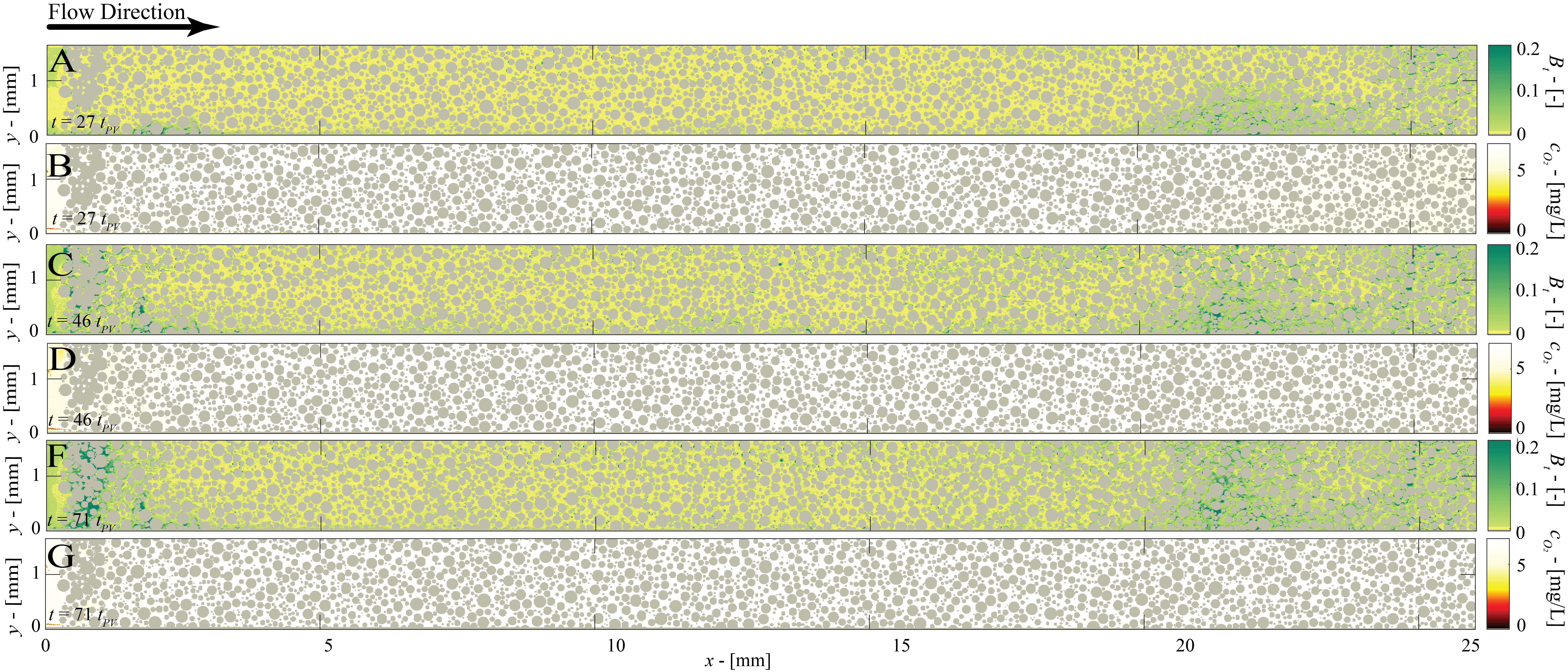}
		\caption{Spatial distribution of $B$ for replicate 2 of the experiment with uncoated chip at $t$ = 27$t_{PV}$ (panel A), $t$ = 46$t_{PV}$ (panel C) and $t$ = 27$t_{PV}$ (panel F) compared to the corresponding $c_{O_2}$ maps ($t$ = 27$t_{PV}$ in panel B, $t$ = 46$t_{PV}$ in panel D, $t$ = 71$t_{PV}$ in panel G)}
		\label{fgr:C10_2}
	\end{figure}
\end{landscape}

\section{References}
[1] M. Moßhammer et al. ACS Sens. 2016, 1, 6, 681–687